\documentclass[aps,pre,twocolumn,nofootinbib,longbibliography,showkeys]{revtex4-2}
\usepackage{cmap}
\usepackage{lmodern}
\usepackage[T1]{fontenc}
\usepackage[utf8]{inputenc}
\input{glyphtounicode}
\usepackage{amsmath,amssymb,bm}
\usepackage{graphicx}
\usepackage[colorlinks=true,linkcolor=blue,citecolor=blue,urlcolor=blue]{hyperref}
\hypersetup{pdftitle={Sleep-Stage Transitions as Landau--Ginzburg Phenomenology},pdfauthor={Alexander Poltorak}}
\graphicspath{{./}}
\begin{document}

\title{A Landau--Ginzburg Phenomenology of Sleep-Stage Transitions}
\author{Alexander Poltorak}
\email{Alex@NeuroLight.co}
\affiliation{NeuroLight, Inc., Pomona, New York 10970, USA}
\date[]{}

\begin{abstract}
Sleep staging provides a reproducible clinical description, but it does not by
itself explain why some boundaries are abrupt while others are graded, or why
transition windows contain instability, synchrony, and apparent state
coexistence. We develop a local Landau--Ginzburg phenomenology in which each
boundary is represented by motion in an effective potential of a spatially
extended, noisy, dissipative neural field. A latent cortical-ordering
coordinate \emph{$\phi$} is inferred from prespecified EEG/PSG observables
through a measurement model designed to avoid circularity. The canonical
boundaries are treated separately. Existing data support a fold-like loss of
wake stability at sleep onset; whether that fold lies on a globally bistable
cusp with hysteresis remains open. N1-to-N2 and N2-to-N3 are posed as
continuous-like ordering crossovers, NREM-to-REM as a candidate
first-order-like desynchronizing switch, and a possible within-N3 mixed or
tricritical-like regime as an explicitly speculative hypothesis. The Ginzburg
term adds spatial predictions---growth of correlation length and
local-to-global recruitment---that are absent from scalar sleep-onset models.
We specify the evidence needed to distinguish bifurcation, coexistence,
noise-driven escape, smooth crossover, and scoring-induced discontinuity.
Illustrative time-dependent Ginzburg--Landau simulations reproduce the
proposed signature classes. A synthetic classification experiment partially
distinguished six archetypes (cross-validated accuracy 0.49 $\pm$ 0.005;
balanced baseline 0.17), with little change under a noise-regime shift. These
analyses establish the internal consistency and testability of the framework,
not the proposed taxonomy in human sleep. Transition-centered EEG validation
is required before clinical or neuromodulation applications are pursued.
\end{abstract}

\maketitle

\section{Introduction}\label{sec:introduction}

Sleep architecture is usually represented as a hypnogram: a discrete
trajectory through wakefulness, non-rapid-eye-movement (NREM) stages N1,
N2, and N3, and rapid-eye-movement (REM) sleep. This representation is
clinically powerful because it converts complex polysomnographic signals
into a small set of reproducible labels \cite{rk1968,silber2007,aasm2023}. The same representation also hides a central
mechanistic problem. Some sleep-state boundaries appear switch-like,
with abrupt changes in arousal, muscle tone, cortical activation, and
responsiveness. Other boundaries appear graded, with slow changes in
spectral power, spindles, K-complexes, slow oscillations, and spatial
synchrony. A single descriptive staging framework therefore mixes
transitions that may have different dynamical signatures.

Computational neuroscience has increasingly treated cortical activity as
a collective phenomenon of large recurrent networks poised near critical
or metastable regimes. Evidence for scale-free neuronal avalanches,
long-range temporal correlations, and synchronization transitions has
motivated models in which the brain occupies a region of parameter space
that balances
 propagation
and stability \cite{bak1987,beggsplenz2003,beggs2008,disanto2018,iyer2018,pearlmutter2009}. Sleep
offers a particularly natural domain for this approach. The brain
repeatedly leaves a high-dimensional wake-like regime, enters
progressively more synchronized NREM states, returns to desynchronized
REM, and repeats this cycle several times across the night.

This paper asks whether sleep-stage boundaries can be described as
phase-transition-like reorganizations in a spatially extended, noisy,
nonequilibrium neural field. The model is phenomenological in the
Landau--Ginzburg sense \cite{landau1937,ginzburg1950}. It does not simulate
every synaptic, thalamic, or neuromodulatory process; instead, it seeks a
small set of collective variables that may summarize the local dynamics near a
boundary.

The model is local rather than a generator of whole-night sleep architecture.
At fixed control parameters, relaxational gradient dynamics settle toward a
minimum and cannot produce ultradian cycling, directional probability
currents, or the REM oscillation. Those phenomena require non-gradient drift,
delay, or Hopf components. Homeostatic pressure, circadian drive, and
neuromodulatory tone are therefore treated here as slowly varying external
inputs, and each stage boundary is analyzed through its own local normal form.
This restriction yields tractable, boundary-specific predictions but should
not be mistaken for a complete model of the sleep cycle.

The proposed mechanisms make different empirical predictions. First-order-like
coexistence should produce bistability, bimodality, hysteresis, path
dependence, or flickering between metastable basins. A deterministic fold
should instead be identified by loss of local stability and critical slowing.
Continuous-like ordering should appear as a smooth change in the order
parameter, accompanied by enhanced fluctuations and growth of spatial
correlation. A mixed or tricritical-like regime should show a continuous
precursor followed by a sharper derivative change, loss of metastability, or
local-to-global nucleation. These signatures can be tested in
transition-centered analyses that use 30-s stage labels as temporal priors
rather than as ground-truth transition times.

Phase-transition models of sleep have important precedents. Cortical
mean-field models describe the sleep cycle in phase-transition terms and
predict a first-order SWS-to-REM transition
\cite{steynross2004,steynross2005,steynross2013}. More recently, Li and
colleagues \cite{li2025} reported fold-like sleep-onset dynamics in two
independent cohorts comprising more than 1000 participants, with a distinct
tipping point preceded by critical slowing. Their reduced model contains a
bistable branch structure, but the study did not compare that global structure
against alternative dynamical models or test matched forward and reverse
trajectories. The local fold is therefore well supported; whether the data
require a physiological cusp with coexistence and hysteresis remains open.
Earlier studies treated sleep stages as attractors and searched sleep EEG for
early-warning signals \cite{demooij2020}, showed that scored stages do not map
one-to-one onto whole-brain states \cite{stevner2019}, and recovered the
wake-to-deep-sleep trajectory in a low-dimensional generative embedding \cite{sanzperl2020}.

The present contribution is narrower than the general
claim that sleep can be modeled using phase-transition concepts. It adds (i) a
spatial Ginzburg term, with predictions for correlation-length growth and
local-to-global recruitment; (ii) a boundary-specific taxonomy that does not
force all stage transitions into one mechanism; and (iii) a
circularity-controlled measurement and validation program. The proposed
within-N3 consolidated-SWS subregime is retained as a clearly marked
conjecture.

\section{Conceptual background: sleep as a dissipative dynamical system}\label{sec:conceptual-background-sleep-as-a-dissipative-dynamical-system}

The sleeping brain is a dissipative biological system. It maintains
organized activity far from thermodynamic equilibrium while continuously
exchanging energy, matter, and information with the body. Classical
sleep physiology explains major state changes through the interaction of
neuromodulatory systems, including aminergic arousal nuclei, cholinergic
REM-generating systems, thalamocortical circuits, the ventrolateral
preoptic area, orexinergic stabilization, and homeostatic sleep pressure
\cite{borbely1982,saper2001,saper2005,lu2006,robinson2011}. These mechanisms provide the biological
substrate for state control. A Landau--Ginzburg description provides a
complementary macroscopic language for the resulting collective
dynamics.

In this language, neuromodulators are treated as control parameters.
Adenosine and related homeostatic variables move the system toward
sleep; circadian arousal changes the effective field favoring
wakefulness or sleep; acetylcholine, norepinephrine, serotonin,
histamine, GABAergic VLPO drive, and thalamocortical gain reshape the
potential landscape. The landscape metaphor is not literal energy
minimization in equilibrium thermodynamics. It represents an effective
nonequilibrium potential, or quasipotential, whose minima correspond to
metastable cortical states.

A scored sleep stage is therefore a coarse-grained region in a
high-dimensional state space. Wake, NREM, and REM can be treated as
attractor families generated by interactions among cortical
excitability, thalamocortical resonance, neuromodulatory tone, and
recurrent network topology rather than only as labels for EEG patterns.
The same scored stage can contain local sleep, microarousals, spindles,
K-complexes, and regional slow waves. This heterogeneity is a reason to
analyze transitions directly instead of treating stages as homogeneous
blocks.

\section{Physical intuition: landscapes, stiffness, and critical slowing down}\label{sec:physical-intuition-landscapes-stiffness-and-critical-slowing-down}

The landscape picture can be understood through a simple mechanical
analogy. In this analogy, the current cortical state is represented by a ball in a
valley. The bottom of the valley is the presently stable attractor.
Neural noise, sensory perturbations, autonomic fluctuations, or
spontaneous microarousals push the ball away from the bottom. If the
valley is steep, the restoring force is strong, and the system rapidly
returns to its prior state. If the valley is shallow, the restoring
force is weak and perturbations decay slowly.

This stiffness is formalized by the leading eigenvalue of the linearized
dynamics around a stable fixed point. If a small perturbation x(t) away
from the fixed point obeys

\begin{equation}
\frac{dx}{dt} = \lambda_{eig}x + \eta(t),\quad\quad\lambda_{eig} < 0,
\label{eq:1}
\end{equation}

then the deterministic recovery time is

\begin{equation}
\tau_{rec} = \frac{1}{\left| \lambda_{eig} \right|}.
\label{eq:2}
\end{equation}

As the system approaches a loss of stability, $\lambda_{\mathrm{eig}} \to 0^{-}$,
recovery becomes slow. In the stochastic case, the same loss of
stiffness produces rising variance, increased lag-1 autocorrelation, and
sometimes flickering between nearby metastable states. These are the
standard early-warning signatures of critical slowing down \cite{scheffer2009}. In sleep physiology, this corresponds to the observation
that a subject near sleep onset may drift, fragment, or alternate
between wake-like and sleep-like microstates before a scored transition
appears.

Here, ``phase-transition-like'' is an operational nonequilibrium description;
equilibrium is not assumed. Minima, barriers, bifurcations, correlation
length, susceptibility, discontinuity, and hysteresis are useful only insofar
as they translate into testable EEG and PSG signatures. A transition class
should be assigned only when those signatures survive controls for scoring
rules, finite-size rounding, filtering, and nonstationarity.

\section{Landau--Ginzburg formalism for cortical sleep transitions}\label{sec:landau-ginzburg-formalism-for-cortical-sleep-transitions}

\subsection{Order parameters and control parameters}\label{sec:order-parameters-and-control-parameters}

Let \emph{$\phi$}(\textbf{r},\emph{t}) $\in$ [0,1] denote a latent measure of cortical
ordering at cortical position r and time t. It is not identified with a single
EEG feature. In empirical work, \emph{$\phi$} should be estimated from a
prespecified measurement model whose indicators may include slow-oscillation
dominance, inverse complexity, and spatial synchrony. Aperiodic spectral
slope, fluctuation statistics, and avalanche measures are treated as
diagnostics unless they are explicitly included in that model.

One single-feature indicator is inverse normalized complexity,

\begin{equation}
\phi_{C}\left( \mathbf{r},t \right) \equiv 1 - C_{norm}\left( \mathbf{r},t \right).
\label{eq:3}
\end{equation}

Wakefulness and REM generally have lower cortical ordering, whereas NREM
deepening is associated with increasing slow-wave organization and synchrony.
Because the Landau polynomial is naturally written about a reference state and
permits signed deviations, we define the centered coordinate
\emph{$\psi$}(\textbf{r},\emph{t}) = \emph{$\phi$}(\textbf{r},\emph{t}) $-$ \emph{$\phi$}\textsubscript{0}.
All potentials below are written in \emph{$\psi$}. Negative values indicate
less ordering than the reference state, not negative biological synchrony. In
the special case \emph{$\phi$} = $\phi_{C}$, inverse
complexity alone serves as the empirical order-parameter indicator.

Control parameters are collected into a vector

\begin{widetext}
\begin{equation}
\mathbf{\lambda}(t) = \{ S(t),C_{circ}(t),ACh(t),NE(t),GABA(t),G_{thal}(t),\ldots\}.
\label{eq:4}
\end{equation}
\end{widetext}

Here \emph{S} is homeostatic sleep pressure, \emph{C\textsubscript{circ}} is circadian
drive, ACh is acetylcholinergic tone, NE is noradrenergic tone, GABA
represents sleep-promoting inhibition, and \emph{G\textsubscript{thal}} represents
thalamocortical gain. These variables reshape the effective landscape; none
defines a sleep stage by itself. Where a scalar control is needed, \emph{u} denotes a reaction coordinate along the physiological path through
\emph{$\lambda$}-space.

\subsection{Effective Landau functional}\label{sec:effective-landau-functional-quasi-potential}

For the gradient approximation used here, the centered cortical-ordering field
is governed by the effective Landau functional

\begin{widetext}
\begin{equation}
\mathcal{F\lbrack}\psi\rbrack = \int_{\Omega}^{}d^{d}r\left\lbrack \frac{a\left( \mathbf{\lambda} \right)}{2}\psi^{2} + \frac{b\left( \mathbf{\lambda} \right)}{4}\psi^{4} + \frac{c}{6}\psi^{6} + \frac{\kappa}{2}|\nabla\psi|^{2} - h\left( \mathbf{\lambda} \right)\psi \right\rbrack,\quad\quad \kappa > 0,\ \text{and either}\ c>0\ \text{or}\ (c=0,\ b>0).
\label{eq:5}
\end{equation}
\end{widetext}

The coefficient \emph{a}(\emph{$\lambda$}) sets the curvature at the reference state, while \emph{b}(\emph{$\lambda$}) controls the leading nonlinear transition character.
The sextic coefficient \emph{c} stabilizes large amplitudes when the quartic term is
negative, \emph{$\kappa$} penalizes spatial inhomogeneity, and \emph{h}(\emph{$\lambda$}) biases the field toward one side of the landscape. These
coefficients are phenomenological; they should be inferred from trajectories
or fitted models rather than mapped one-to-one onto neurotransmitters.

No exact biological symmetry underwrites the transformation \emph{$\psi$} $\rightarrow$ $-\psi$. The even expansion is a centered local normal
form, not a claim that cortical ordering has a physical reflection symmetry.
The field \emph{h} is generally nonzero. With \emph{b} \textgreater{} 0 and \emph{h} $\simeq$ 0, changing \emph{a} through zero gives the idealized continuous transition; a finite
field rounds it into a crossover. By contrast, a quartic cusp can retain two
basins and hysteresis over a finite range of bias up to the spinodals.

A cubic term need not be written explicitly in the quartic sector. The translation
\emph{$\psi$} $\rightarrow$ \emph{$\psi$} $-$ \emph{B}/(4\emph{A}) converts a generic quartic \emph{A}\emph{$\psi$}\textsuperscript{4} + \emph{B}\emph{$\psi$}\textsuperscript{3} + \emph{C}\emph{$\psi$}\textsuperscript{2} + \emph{D}\emph{$\psi$} into a depressed quartic
and absorbs the cubic contribution into redefined quadratic and linear
coefficients. The first-order hypotheses below therefore use the quartic cusp as their generic local normal form.

The sextic term in Eq.~(\ref{eq:5}) serves
a different purpose. It stabilizes the potential when \emph{b} \textless{} 0 and contains the fine-tuned tricritical limit. An exact mean-field tricritical point requires \emph{a} = \emph{b} = \emph{h} = 0 with \emph{c} \textgreater{} 0 and an approximate
symmetry that is not derived here. Accordingly, the within-N3 proposal is
framed as a mixed or tricritical-like crossover, not as a claim that human
sleep is tuned to an exact tricritical point.

The corresponding stochastic dynamics are written as a time-dependent Ginzburg--Landau equation,

\begin{equation}
\frac{\partial\psi\left( \mathbf{r},t \right)}{\partial t} = - \Gamma\frac{\delta\mathcal{F}}{\delta\psi\left( \mathbf{r},t \right)} + \eta\left( \mathbf{r},t \right),
\label{eq:6}
\end{equation}

Here \emph{$\Gamma$} is a relaxation coefficient and \emph{$\eta$}(\textbf{r},\emph{t})
represents neural, sensory, and autonomic fluctuations. Measurement and
scoring errors belong to the observation model, not to the latent-field noise.
Additive white noise is used only in the illustrative simulations; empirical
data may require colored, multiplicative, or state-dependent noise. No
equilibrium fluctuation-dissipation relation is assumed. The functional
derivative is

\begin{equation}
\frac{\delta\mathcal{F}}{\delta\psi} = a\left( \mathbf{\lambda} \right)\psi + b\left( \mathbf{\lambda} \right)\psi^{3} + c\psi^{5} - \kappa\nabla^{2}\psi - h\left( \mathbf{\lambda} \right).
\label{eq:7}
\end{equation}

Therefore, the explicit neural-field equation is

\begin{equation}
\frac{\partial\psi}{\partial t} = - \Gamma\left\lbrack a\psi + b\psi^{3} + c\psi^{5} - \kappa\nabla^{2}\psi - h \right\rbrack + \eta.
\label{eq:8}
\end{equation}

Linearizing Eq.~(\ref{eq:8}) near a spatially uniform stable state
\emph{$\psi$\textsubscript{eq}} gives

\begin{equation}
\frac{\partial\delta\psi}{\partial t} = - \Gamma\left\lbrack a + 3b\psi_{eq}^{2} + 5c\psi_{eq}^{4} \right\rbrack\delta\psi + \Gamma\kappa\nabla^{2}\delta\psi + \eta.
\label{eq:9}
\end{equation}

The bracketed term is the local curvature of the potential. As it
approaches zero, the state loses stiffness, recovery slows, variance
rises, and the system becomes susceptible to noise-driven switching. In
the dynamical equation the Ginzburg term contributes positive diffusion,
+\emph{$\Gamma$} \emph{$\kappa$} $\nabla$\textsuperscript{2} \emph{$\psi$}, which smooths local fluctuations when
\emph{$\kappa$} \textgreater{} 0 rather than producing anti-diffusion.
\begin{table*}[t]
\caption{Operational interpretation of Landau--Ginzburg coefficients.}
\label{tab:box1}
\footnotesize
\setlength{\tabcolsep}{4pt}
\noindent\begin{tabular}{@{}p{0.20\textwidth}p{0.25\textwidth}p{0.25\textwidth}p{0.20\textwidth}@{}}
\hline\hline
\textbf{Coefficient} & \textbf{Mathematical role} & \textbf{Biological interpretation} & \textbf{Empirical proxy} \\
\hline
\emph{$a$} & quadratic coefficient; curvature contribution at the reference state & contribution to basin stiffness; exact local stiffness is $V''(\psi_{eq})$ & recovery time, variance, lag-1 autocorrelation \\
\emph{$b$} & nonlinear transition character & tendency toward smooth ordering vs bistability & model comparison, bimodality, path dependence \\
\emph{$c$} & sextic stabilization & high-amplitude stabilization/saturation of ordering & saturation of slow-wave dominance at high \emph{$\phi$} \\
\emph{$\kappa$} & spatial coupling/gradient cost & local-to-global recruitment of cortical domains & correlation length (identifies combinations such as $\kappa/V''$, not $\kappa$ alone), traveling-wave coherence \\
\emph{$h$} & biasing field & arousal, REM, circadian, or sensory asymmetry & EOG/EMG/autonomic state, circadian phase, stimulus context \\
\emph{$\Gamma$} & relaxation rate & speed of return after perturbation & microarousal recovery, perturbation-response slope \\
\emph{$\eta$} & noise term & neural, autonomic, and sensory fluctuations & residual variability, flickering, noise-color estimates \\
\hline\hline
\end{tabular}
\end{table*}

\begin{figure*}[t]
\centering
\includegraphics[width=\textwidth]{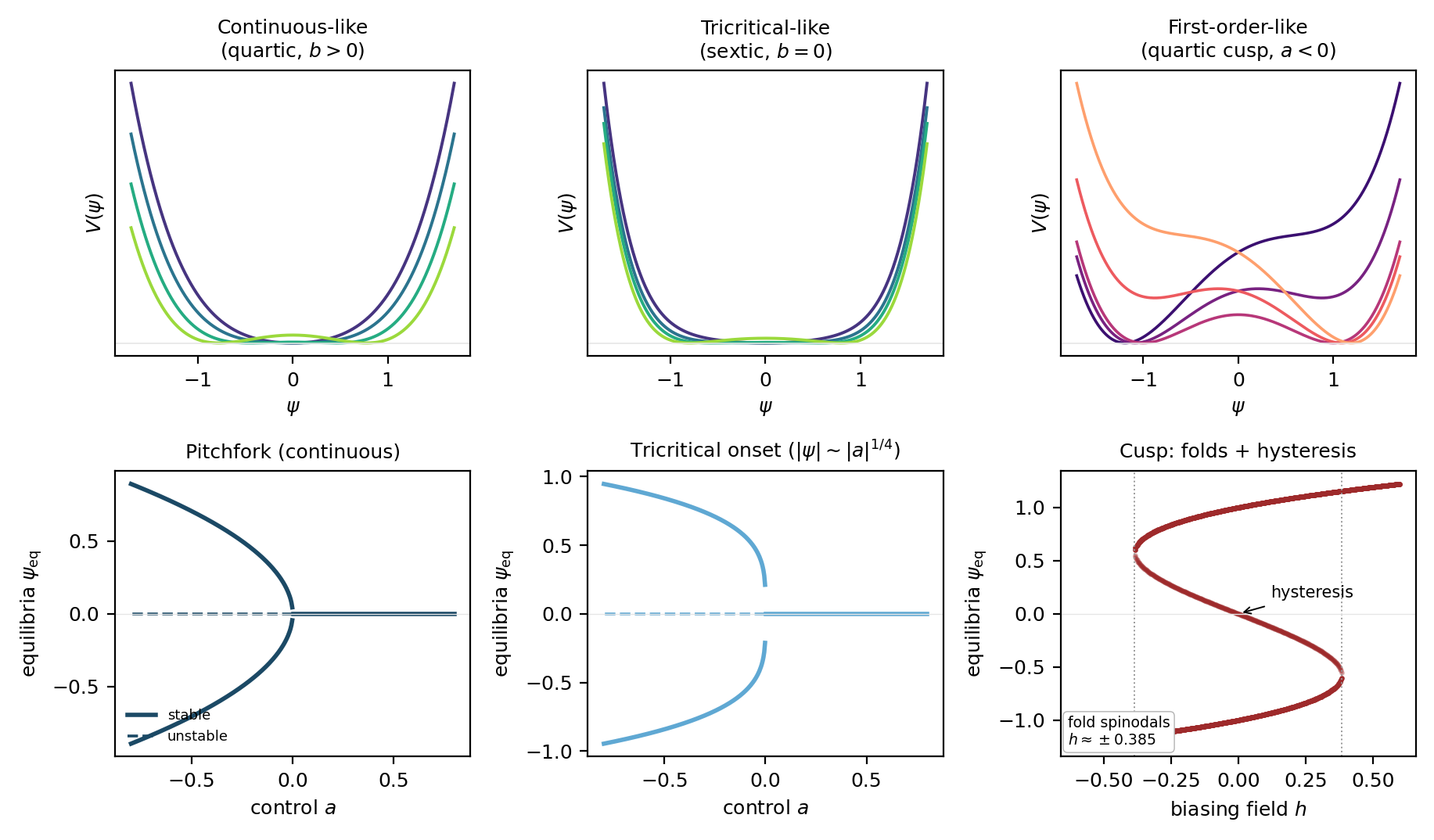}
\caption{Transition order is set by the control path, not by the static appearance of a single potential. \emph{Top row:} families of the effective Landau potential $V(\psi)$ as a control parameter is swept, for the continuous-like (quartic, $b>0$, $h\approx 0$), tricritical-like ($b\to 0$), and first-order-like (quartic cusp, $a<0$) cases; the horizontal axis is the signed Landau coordinate \emph{$\psi$}, not negative biological synchrony. \emph{Bottom row:} the corresponding bifurcation diagrams of the equilibria along the swept control parameter, with solid curves marking stable branches and dashed curves unstable branches. The first-order-like column shows the fold spinodals (\emph{h} $\approx$ $\pm$0.385 for the normalized cusp with $a=-1$ and unit quartic coefficient) at which a branch is lost, and the resulting hysteresis loop under a swept field; these features, and the fold--hysteresis relationship, not the bare double well, are what the empirical tests target.}
\label{fig:pot}
\end{figure*}

\subsection{Relation to neural mass and criticality models}\label{sec:relation-to-neural-mass-and-criticality-models}

The proposed phenomenology is compatible with established neural mass
models. The Steyn-Ross cortical mean-field model treats the cortex as a
spatially extended excitatory-inhibitory medium and explicitly models
sleep cycling in phase-transition terms, including an SWS-to-REM
transition from a low-firing coherent state to a high-firing
desynchronized state \cite{steynross2005}. The Phillips-Robinson
sleep model describes subcortical sleep-wake regulation through
interacting mean soma voltages and homeostatic drive, producing
flip-flop bistability between wake-promoting and sleep-promoting
populations \cite{robinson2011}. These prior models are the closest
precedents for the present framework; the novelty here lies in assigning
different operational transition classes to individual stage boundaries
and in specifying EEG tests for each boundary.

Criticality models provide a second foundation, though avalanche measures
require care. If \emph{N}\textsubscript{t} = \emph{n} \textgreater{} 0 events are active in
one time bin, the conditional branching ratio is

\begin{equation}
\sigma = \frac{\mathbb{E}\left\lbrack N_{t + 1} \mid N_{t} \right\rbrack}{N_{t}}.
\label{eq:10}
\end{equation}

Wakefulness and REM are often associated with richer long-range temporal
structure and higher effective complexity, while NREM sleep exhibits
stronger slow oscillations and reduced long-range temporal correlations
\cite{meisel2017,priesemann2013}. Di Santo and colleagues~\cite{disanto2018} showed that a Landau--Ginzburg theory of the cortex can generate down
states, asynchronous activity, synchronized phases, and scale-free
avalanches near the edge of synchronization. The same
Landau--Ginzburg cortical description has been analyzed at the edge of a
discontinuous transition, where \emph{self-organized bistability} rather
than self-organized criticality governs the collective
dynamics~\cite{buendia2020}, a distinction that mirrors the first-order
versus continuous-like separation maintained in the taxonomy developed
here. This supports using
the same formal language for sleep-stage dynamics, but avalanche
branching should be treated as exploratory in sparse scalp PSG because
estimates are sensitive to thresholding, subsampling, referencing, and
filtering.

The present model does not require every sleep state to be literally at
a thermodynamic critical point. The claim is more precise: different
sleep-stage boundaries may instantiate different forms of dynamical
reorganization, and their signatures can be distinguished empirically
with tools adapted from nonequilibrium statistical mechanics.

\subsection{Distinguishing dynamical mechanisms}\label{sec:distinguishing-dynamical-mechanisms}

A recurring hazard in this area is to treat ``first-order-like,''
``saddle-node-like,'' ``critical slowing,'' ``attractor switching,'' and
``scoring-induced discontinuity'' as one interchangeable category. They are not,
and the framework is only falsifiable if they are kept apart. At least
five mechanisms can generate a transition that looks abrupt or graded in
scored data, and each predicts different evidence. A first-order
(coexistence) transition occurs when two competing minima of the
effective potential exchange global stability; its hallmarks are
bimodality, hysteresis, and path dependence that survive when control
parameters are matched. A deterministic bifurcation of a finite neural
system, such as a fold, is a loss of local stability of a fixed point;
its hallmark is critical slowing, a rise in recovery time, variance, and
lag-1 autocorrelation as the tipping point is approached, as reported
empirically for sleep onset \cite{li2025}. Metastable stochastic
switching is noise-driven escape across a barrier before any
deterministic instability is reached; it produces dwell-time
distributions and flickering rather than a fixed threshold. A smooth
crossover is a continuous, reversible change with no coexistence and no
true singularity, which nonzero fields and finite size make the default
expectation for the continuous boundaries. Finally, a scoring-induced
discontinuity is an artifact of applying 30-second categorical rules to
continuous data, and can manufacture apparent jumps where the underlying
trajectory is smooth. Critically, a rise in autocorrelation alone
establishes none of these: it is consistent with an approaching
bifurcation, with barrier softening, and with mere spectral change. The
taxonomy below states, for each archetype, the supportive and the distinguishing evidence; the
operational tests are specified in Sec.~\ref{sec:confounds-and-identifiability}.

\begin{table*}[t]
\caption{Dynamical properties, generator regimes, and distinguishing evidence for the candidate transition archetypes. Bimodality and jumps are expected or supportive signatures, not literally necessary in every finite, noisy realization; a single physiological window may combine several rows.}
\label{tab:mech}
\footnotesize
\setlength{\tabcolsep}{4pt}
\noindent\begin{tabular}{@{}p{0.20\textwidth}p{0.25\textwidth}p{0.25\textwidth}p{0.20\textwidth}@{}}
\hline\hline
\textbf{Archetype} & \textbf{Working definition} & \textbf{Expected or supportive evidence} & \textbf{Evidence that distinguishes it} \\
\hline
First-order (coexistence) transition & Two potential minima exchange global stability & Bimodality plus a discontinuous jump in \emph{$\phi$} & Hysteresis and path dependence that persist at matched control values \\
Deterministic bifurcation (e.g.~fold) & Local annihilation of a stable and an unstable branch & Reproducible tipping point; slowing recovery and rising autocorrelation under adequate detrending & Local branch termination; this does not by itself determine whether another stable branch exists globally \\
Metastable stochastic switching & Noise-driven escape before deterministic instability & Broad dwell-time distributions; flickering & Escape statistics that scale with noise, not a fixed threshold \\
Smooth crossover & Continuous reversible change, no singularity & Monotonic \emph{$\phi$} with no bimodality and no jump & Reversibility and absence of hysteresis; rounding scales with the biasing field \\
Scoring-induced discontinuity & Artifact of 30-second categorical scoring & Jump present in labels but not in continuous features & Disappearance under jittered or re-estimated boundaries \\
Continuous (second-order-like) ordering & Order parameter grows continuously from zero as $a \to 0^{-}$ & Enhanced variance and lag-1 autocorrelation with a growing correlation length & Smooth, reversible growth of \emph{$\phi$} with no jump or bimodality \\
Tricritical limiting behavior & Continuous ordering with exponent $1/4$ at $a=b=h=0$ ($c>0$) & Continuous, reversible growth with mean-field exponent $1/4$ & Steeper onset than the ordinary continuous case, still with no jump or bimodality \\
Mixed continuous-plus-switch & Continuous precursor followed by a distinct switch or nucleation (near or past $b\to 0$) & Continuous run-up plus a derivative change, late bimodality, or local-to-global nucleation & A distinguishable substate beyond a rounded crossover \\
\hline\hline
\end{tabular}
\end{table*}

The corresponding local normal forms make these distinctions concrete. A fold (saddle-node) is $\dot\psi$ = r $-$
$\psi$\textsuperscript{2}, with a stable and an unstable branch that annihilate at r = 0; this
is the form supported empirically at sleep onset. A symmetric continuous
transition is the supercritical pitchfork $\dot\psi$ = $-$a \emph{$\psi$} $-$ \emph{b} $\psi$\textsuperscript{3}
with \emph{b} \textgreater{} 0, whose ordered branch grows as $\sqrt{-a}$; a
nonzero field \emph{h} turns this into an imperfect (rounded)
bifurcation, which is why the continuous cases are described as
crossovers. First-order coexistence is the cusp obtained from the
quartic-with-field potential, giving two stable branches separated by a
fold on each side and a hysteresis loop under a swept field. Tricritical
behavior requires the sextic term together with the approximate symmetry
and the extra tuning noted above. Metastable stochastic switching is not
a bifurcation at all: at fixed control, the system escapes over a barrier
at a noise-dependent rate given by Kramers' law, producing flips without
a deterministic threshold. Throughout the taxonomy of Sec.~\ref{sec:proposed-taxonomy-of-sleep-stage-transitions}, the
label attached to a boundary should be read as a hypothesis about which
of these dynamical properties or archetypes is expressed, testable by the corresponding row of
this table, and not as a claim that the boundary is a literal
equilibrium phase transition.

\subsection{Measurement model for the latent order parameter}\label{sec:measurement-model-for-the-latent-order-parameter}

Because \emph{$\phi$} is latent, any claim about
transition mechanism depends on how \emph{$\phi$} is estimated from recorded
signals. We therefore specify a measurement model rather than an ad hoc
composite. Let \(\mathbf{y}(t)\) be a vector of prespecified,
standardized EEG/PSG features, for example slow-oscillation dominance,
inverse Lempel-Ziv complexity, and a spatial-synchrony index, and let

\begin{equation}
\mathbf{y}(t) = \mathbf{\Lambda}\,\phi(t) + \mathbf{\varepsilon}(t),
\label{eq:11}
\end{equation}

with loading vector \(\mathbf{\Lambda}\), measurement noise
\(\mathbf{\varepsilon}(t)\), and \emph{$\phi$} recovered as the latent factor
of a state-space or latent-variable model. Equation~(\ref{eq:11}) is
written for a spatially aggregated order parameter $\phi(t)$. For scalp
EEG, the channelwise generalization must account for the lead field,
$y_i(t) = \int_{\Omega} L_i(\mathbf{r})\,\Lambda(\mathbf{r})\,\phi(\mathbf{r},t)\,d\mathbf{r} + \varepsilon_i(t)$,
where $L_i$ is the lead-field kernel; the local form
$y_i(t) = \Lambda_i\,\phi(\mathbf{r}_i,t) + \varepsilon_i(t)$ is appropriate only
after source reconstruction or for sufficiently local intracranial recordings. Because a
latent factor is defined only up to sign and scale, we anchor it by
fixing its orientation (positive toward greater slow-wave ordering) and
by imposing one identification constraint---either fixing a single marker
loading to unity or fixing the latent variance---so that the recovered
coordinate is identified. Because the biological ordering is bounded, we
prefer the marker-loading constraint so that the latent coordinate retains its
intended biological scale rather than being standardized to unit variance; for
a coordinate normalized to $0\le\phi\le1$, unit variance is in any case
unattainable ($\mathrm{Var}(\phi)\le 1/4$). The remaining loadings and the latent
variance are then estimated. A linear Gaussian factor model does not itself
enforce boundedness; where boundedness is essential rather than interpretive,
empirical implementations should use a bounded-state or transformed latent model
(e.g., a logistic link). Three requirements make this
more than a relabeling. First, the loadings, normalization, and feature
set must be prespecified and, ideally, preregistered, and the recovered
transition mechanism must be shown to be robust to reasonable
alternative feature sets and normalizations. Second, measurement
invariance must be tested across subjects, cycles, ages, and montages,
because a composite whose loadings drift across these strata can
manufacture or erase bimodality and derivative changes. Third, and most
important for falsifiability, \emph{$\phi$} must not be constructed from the
same features that define the boundary under test: estimating \emph{$\phi$}
from slow-wave dominance and then using slow-wave-defined N3 boundaries
to test for a within-N3 transition is circular. Where a boundary is
defined by a feature, that feature must be excluded from, or explicitly
separated in, the construction of \emph{$\phi$} for that test. In this paper, the measurement model is specified but not fitted; fitting and
invariance testing are part of the validation program of Secs.~\ref{sec:operational-eeg-metrics-and-falsifiable-predictions} and \ref{sec:proposed-methods-for-empirical-validation}.

\section{Mapping classical sleep models into the phase-transition framework}\label{sec:mapping-classical-sleep-models-into-the-phase-transition-framework}

\subsection{Borb\'ely two-process regulation as parameter drift}\label{sec:borbely-two-process-regulation-as-parameter-drift}

Borb\'ely's two-process model describes sleep propensity as the
interaction between Process S, a homeostatic variable that increases
during wake and declines during sleep, and Process C, a circadian
variable that gates the timing of sleep and wake \cite{borbely1982,borbely2016}. In the present formalism, Process S and Process C define
a trajectory through control-parameter space:

\begin{equation}
\mathbf{\lambda}(t) = \mathbf{\lambda}\left\lbrack S(t),C_{circ}(t),\ldots \right\rbrack.
\label{eq:12}
\end{equation}

They do not by themselves specify the order of a transition. A smooth
change in Process S can move the system across a bifurcation, producing
an abrupt state change when an attractor disappears or when a competing
basin becomes dominant.

This distinction resolves an apparent paradox of sleep onset. Subjective
sleepiness and homeostatic pressure may evolve gradually across the
evening, while actual loss of wake stability can occur sharply. In
landscape terms, Process S slowly deforms the wake basin. The wake state
can remain locally stable until the barrier becomes low enough for
noise, behavioral context, eye closure, reduced sensory input, and
circadian phase to trigger a transition.

\subsection{Flip-flop circuitry as bistability in the effective potential}\label{sec:flip-flop-circuitry-as-bistability-in-the-effective-potential}

The flip-flop switch model describes mutual inhibition between
sleep-promoting VLPO neurons and wake-promoting monoaminergic and
orexin-stabilized arousal systems \cite{saper2001,saper2005}. Reciprocal inhibition naturally produces bistability, which
corresponds in the Landau description to two local minima separated by
an unstable barrier. The circuit model identifies a biological
implementation of the potential landscape. The phenomenological model
classifies the macroscopic transition that the circuit implements.

A minimal bistable reduction can be written as

\begin{equation}
\begin{aligned}
\frac{d\psi}{dt} &= - \Gamma\frac{dV\left( \psi;\mathbf{\lambda} \right)}{d\psi} + \eta(t),\\
V(\psi) &= \frac{a}{2}\psi^{2} + \frac{b}{4}\psi^{4} + \frac{c}{6}\psi^{6} - h\psi.
\end{aligned}
\label{eq:13}
\end{equation}

When \emph{V} has two minima, a flip-flop circuit can be interpreted as the
neural implementation of the same bistable landscape. The same circuit
motif can participate in transitions with different apparent orders
depending on the parameter regime. A heavily biased, barrier-dominated
landscape gives switch-like transitions. A shallow, spatially extended
landscape with a positive quartic coefficient supports continuous growth
of synchrony and correlation length. The question becomes empirical: at
each stage boundary, which landscape signature is observed?

\section{Proposed taxonomy of sleep-stage transitions}\label{sec:proposed-taxonomy-of-sleep-stage-transitions}

\begin{figure*}[t]
\centering
\includegraphics[width=\textwidth]{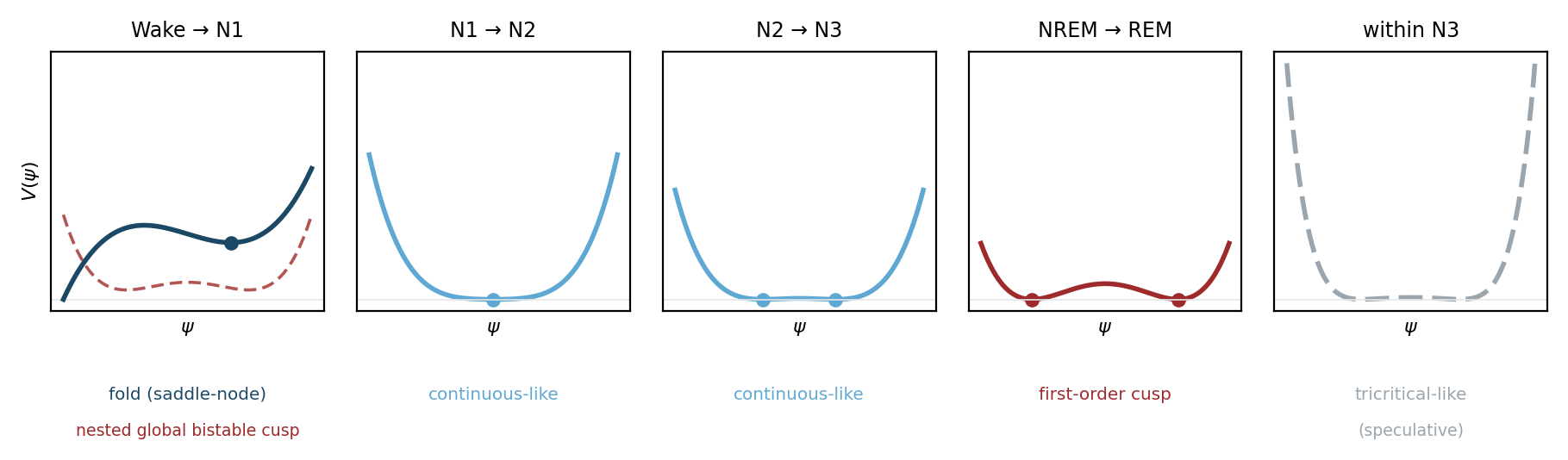}
\caption{The boundary-specific taxonomy, drawn as a set of independent local transition windows rather than a single night-long trajectory, consistent with the local scope of the framework. Each panel shows the schematic effective potential and the hypothesized dynamical mechanism at one boundary: an empirically supported fold for wake-to-N1, with the nested global bistable-cusp hypothesis (whose spinodal is that fold) drawn faintly in the same panel; a first-order cusp for NREM-to-REM; continuous-like crossovers for N1-to-N2 and N2-to-N3; and a speculative tricritical-like subregime within N3. The horizontal axis is the signed Landau coordinate; the panels are conceptual and imply no validated numeric scale for \emph{$\psi$} and no global sleep-cycle trajectory.}
\label{fig:tax}
\end{figure*}

\subsection{Wakefulness to N1: a supported fold and a conditional cusp hypothesis}\label{sec:wakefulness-to-n1-an-empirically-supported-fold-with-a-competing-coexistence-hypothesis}

Sleep onset is the best-supported boundary in the taxonomy. Li et al.
\cite{li2025} collapsed the approach to sleep onto a low-dimensional
feature-space coordinate and found dynamics consistent with a fold bifurcation
in two cohorts totaling more than 1000 participants. Variance and
autocorrelation rose before a tipping point several minutes before
conventional sleep onset. Their analysis did not formally compare the fold
model with alternative dynamical families. Moreover, although their chosen
reduced model contains stable and unstable branches, the data were not used to
test matched forward and reverse trajectories. The empirical result therefore
supports local loss of wake stability; it does not by itself establish physiological hysteresis.

Hu et al. \cite{hu2026} provide complementary
evidence from a smaller sample. Their stochastic double-well model was fitted
directly to sleep-onset EEG and reproduced noise-driven switching between
wake-like and sleep-like states before final commitment to sleep. The fitted
drift and noise parameters correlated with subjective sleepiness. Nineteen of
37 enrolled participants contributed usable data, and the study did not
compare the bistable model with a pure-fold alternative. It is therefore
evidence for the plausibility of bistable sleep-onset dynamics, not a definitive model selection.

We distinguish the local and global questions.
Locally, wake stability may terminate at a fold. Globally, that fold may be a
spinodal of a cusp in which wake and a sleep-like state coexist over a finite
control range. The Ginzburg extension adds a spatial question: does the
approach to the fold involve growth of cortical correlation length and
local-to-global recruitment? N1 is heterogeneous and strongly affected by eye
closure, posture, sensory disengagement, and behavioral context, so these
variables must be modeled rather than folded into the transition mechanism.

A minimal sleep-onset potential, written as the global quartic cusp
whose local expansion near either spinodal reduces to the fold normal
form, is

\begin{widetext}
\begin{equation}
V_{W \rightarrow N1}(\psi) = \frac{a_{W}\left( S,C_{circ} \right)}{2}\psi^{2} + \frac{1}{4}\psi^{4} - h_{W}\left( S,C_{circ},NE \right)\psi,\quad\quad a_{W} < 0\ \text{(quartic cusp)}.
\label{eq:14}
\end{equation}
\end{widetext}

The stationary states satisfy

\begin{equation}
\frac{dV_{W \rightarrow N1}}{d\psi} = a_{W}\psi + \psi^{3} - h_{W} = 0.
\label{eq:15}
\end{equation}

The stability condition is

\begin{equation}
\frac{d^{2}V_{W \rightarrow N1}}{d\psi^{2}} = a_{W} + 3\psi^{2} > 0.
\label{eq:16}
\end{equation}

If the fold is embedded in a bistable cusp, wake and sleep attractors
remain simultaneously stable over a finite control region, and sleep
onset should then show transition-path dependence: the threshold for
wake-to-sleep entry need not equal the threshold for sleep-to-wake
re-entry. This provides a concrete way to test hysteresis using repeated
transitions, microarousals, and nap protocols. Critical slowing down is
expected specifically near loss of local stability, such as a
saddle-node or spinodal-like boundary; it should not be treated as a
generic property of all first-order-like switching. The abruptness
observed at the behavioral level can coexist with a gradual
neurochemical run-up because the control parameter changes continuously
while the order parameter changes discontinuously when local stability
is lost.

\subsection{N1-to-N2: continuous thalamocortical ordering}\label{sec:n1-to-n2-continuous-thalamocortical-ordering}

N2 is not simply an intermediate scalar level of ordering between N1 and N3.
It is defined by discrete, event-like graphoelements, spindles and
K-complexes, that reflect specific thalamocortical and cortical down-state
dynamics rather than a uniform increase in a single synchronization amplitude.
A K-complex resembles an isolated cortical down-state more than a small
increment of \emph{$\phi$}. The continuous-ordering description therefore
applies, at most, to a coarse-grained time-averaged coordinate. At finer
resolution the boundary is partly event driven, and an event-based model may
outperform the scalar reduction.

The N1-to-N2 boundary is modeled as a plausible but under-tested
continuous-like ordering crossover. N2 is defined by sleep spindles and
K-complexes, which reflect thalamocortical coordination, cortical
down-state dynamics, and a more stable NREM regime \cite{cash2009,degennaro2003,fernandez2020}. In the Landau
language, local clusters of coordinated activity may grow as arousal
tone declines and thalamocortical gain becomes more favorable to NREM
oscillations. However, the first scored appearance of spindles or
K-complexes can itself create an apparent boundary, so empirical tests
must analyze continuous spindle density, sigma power, K-complex
probability, and complexity around the transition rather than treating
the scored label as the transition itself.

The local potential for a continuous N1-to-N2 transition can be written
as

\begin{equation}
V_{N1 \rightarrow N2}(\psi) = \frac{1}{2}a_{12}(u)\psi^{2} + \frac{1}{4}b_{12}\psi^{4},\quad\quad b_{12} > 0,
\label{eq:17}
\end{equation}

with

\begin{equation}
a_{12}(u) = \alpha_{12}\,u.
\label{eq:18}
\end{equation}

The equilibrium solution is

\begin{equation}
\psi_{eq} = 0\quad\text{for }a_{12} > 0,
\label{eq:19}
\end{equation}

and

\begin{equation}
\psi_{eq} = \pm\sqrt{\frac{- a_{12}}{b_{12}}}\quad\text{for }a_{12} < 0.
\label{eq:20}
\end{equation}

Thus the order parameter grows continuously from zero as the system
crosses the idealized critical boundary. In real PSG, nonzero biasing
fields and scoring rules may round this into a continuous crossover
rather than a sharp second-order transition. The expected empirical
pattern is a smooth decline in complexity and high-frequency activation,
a gradual increase in spindle-band organization, and enhanced fluctuation
variance (and, where perturbation data are available, increased response
susceptibility) near the boundary. The emergence of
spindle-band organization may be interpreted as one marker of increasing
thalamocortical order, not as proof of symmetry breaking by itself. The
relevant empirical question is whether N1-to-N2 shows a jump in order
parameters or a continuous trajectory with enhanced variance and
spatial/temporal coherence.

\subsection{N2-to-N3: continuous growth of slow-wave synchronization}\label{sec:n2-to-n3-continuous-growth-of-slow-wave-synchronization}

The N2-to-N3 transition is the most plausible candidate in this
framework for a continuous ordering process. Slow-wave activity
increases, cortical responsiveness decreases, high-frequency activity
diminishes, and traveling slow waves become more prominent. The order
parameter \emph{$\phi$} should rise gradually as local cortical slow
oscillations recruit larger territories. The gradient term is essential
here because N3 is spatially structured: slow waves often begin locally
and propagate, with regional variability in timing and amplitude \cite{huber2004,massimini2004,nir2011,vyazovskiy2011,achermann1997}.

A spatial model for the N2-to-N3 transition is

\begin{widetext}
\begin{equation}
\mathcal{F}_{N2 \rightarrow N3}\lbrack\psi\rbrack = \int_{\Omega}^{}d^{d}r\left\lbrack \frac{a_{23}(\lambda)}{2}\psi^{2} + \frac{b_{23}}{4}\psi^{4} + \frac{\kappa_{23}}{2}|\nabla\psi|^{2} \right\rbrack,\quad\quad b_{23} > 0.
\label{eq:21}
\end{equation}
\end{widetext}

The spatial correlation length is predicted to scale as

\begin{equation}
\xi_{23} = \sqrt{\frac{\kappa_{23}}{V''(\psi_{eq})}},\qquad V''(\psi_{eq}) = a_{23} + 3 b_{23}\psi_{eq}^{2}.
\label{eq:22}
\end{equation}

On the disordered side ($\psi_{eq}\approx 0$, $a_{23}>0$), this reduces to
$\xi_{23}=\sqrt{\kappa_{23}/|a_{23}|}$; on the ordered side
($\psi_{eq}^{2}=-a_{23}/b_{23}$) the curvature is $2|a_{23}|$, so
$\xi_{23}=\sqrt{\dfrac{\kappa_{23}}{2|a_{23}|}}$. Both sides exhibit the
mean-field/Gaussian scaling $\xi_{23}\propto|a_{23}|^{-1/2}$, with different amplitudes.
As $a_{23} \to 0$, $\xi_{23}$ increases. In EEG terms, this corresponds
to a transition from local slow-wave events toward broader cortical
recruitment. The static susceptibility, defined as the linear response
$\chi_{23} = \partial\langle\psi\rangle/\partial h$ of the order parameter
to the biasing field, also increases near the boundary:

\begin{equation}
\chi_{23} = \frac{1}{V''(\psi_{eq})} \propto \frac{1}{\left| a_{23} \right|}.
\label{eq:23}
\end{equation}

This is a response quantity; absent an equilibrium
fluctuation-dissipation relation, it is not equivalent to the
spontaneous fluctuation variance, as cautioned below.

The predicted signatures are smooth changes in Lempel-Ziv complexity
\cite{lempel1976}, multiscale entropy \cite{costa2002},
aperiodic slope, phase synchrony, and slow-oscillation power. A variance
peak or increase in lagged autocorrelation around the transition would
support critical-like fluctuations. Absence of bimodality would argue
against a first-order account of the N2-to-N3 boundary at the scoring
level, but any such conclusion must control for the AASM slow-wave
threshold that defines N3.

\subsection{A speculative within-N3 mixed-transition subregime (tricritical limiting case)}\label{sec:a-speculative-within-n3-mixed-transition-subregime-tricritical-limiting-case}

\textbf{Speculative hypothesis.} Before assigning a transition order within
N3, consolidated SWS must first be shown to be a reproducible substate rather
than the upper tail of a continuous sleep-depth variable. A test based on the
same slow-wave features used to define the substate would be circular, and R\&K
Stage 3-to-Stage 4 labels do not solve that problem because they are
themselves based on high-amplitude slow-wave content. The first analysis
should therefore compare preregistered one-state continuous, two-state
switching, and continuous-plus-switch models while excluding the defining
feature from the construction of \emph{$\phi$}
(Sec.~\ref{sec:measurement-model-for-the-latent-order-parameter}).

N3 is the deepest official AASM NREM stage, but it contains residual spindles,
K-complexes, local slow waves, and varying degrees of spatial coherence. The
proposed consolidated-SWS subregime is defined operationally by sustained
slow-oscillation dominance, high spatial coherence, low complexity, and
repeated up/down-state organization. It should be described as consolidated
SWS or an N4-like substate, not as an official AASM stage. R\&K Stage
3-to-Stage 4 transitions provide only an operational approximation.

The working hypothesis is that early NREM deepening is continuous, whereas the
final descent into highly consolidated SWS may show mixed-transition
signatures. A tricritical Landau point is one compact mathematical
idealization of this possibility, not a claim that human N3 is
generically tuned to an exact tricritical point. In Landau terms, the
quartic coefficient changes with control parameters:

\begin{equation}
b_{SWS}(u) = \beta_{0}\left( u - u_{tri} \right).
\label{eq:24}
\end{equation}

When $b_{SWS}$ \textgreater{} 0, the transition behaves as an ordinary
continuous ordering process. Along the approximately symmetric
$h \approx 0$ sector, $b_{SWS} = 0$ marks the boundary of the
sextic-dominated regime,

\begin{equation}
b_{SWS} = 0,\quad\quad c_{SWS} > 0,
\label{eq:25}
\end{equation}

while the mean-field tricritical point itself requires
$a_{SWS} = b_{SWS} = h = 0$ with $c_{SWS} > 0$.

The tricritical potential is

\begin{equation}
V_{tri}(\psi) = \frac{a_{SWS}}{2}\psi^{2} + \frac{c_{SWS}}{6}\psi^{6}.
\label{eq:26}
\end{equation}

For $a_{SWS}$ \textless{} 0, the ordered solution scales as

\begin{equation}
|\psi_{eq}| \sim \left( \frac{- a_{SWS}}{c_{SWS}} \right)^{1/4},
\label{eq:27}
\end{equation}

rather than the ordinary second-order scaling

\begin{equation}
|\psi_{eq}| \sim \left( \frac{- a}{b} \right)^{1/2}.
\label{eq:28}
\end{equation}

On the $b_{SWS} < 0$ side of the tricritical manifold, the sixth-order
term stabilizes the potential and permits first-order-like coexistence:

\begin{widetext}
\begin{equation}
V_{SWS}(\psi) = \frac{a_{SWS}}{2}\psi^{2} + \frac{b_{SWS}}{4}\psi^{4} + \frac{c_{SWS}}{6}\psi^{6},\quad\quad b_{SWS} < 0,\ c_{SWS} > 0.
\label{eq:29}
\end{equation}
\end{widetext}

For the symmetric case \emph{h}=0, coexistence of the \emph{$\psi$} = 0 and \emph{$\psi$}
$\neq$ 0 minima occurs at

\begin{equation}
a_{coex} = \frac{3b_{SWS}^{2}}{16c_{SWS}},
\label{eq:30}
\end{equation}

with the discontinuous jump to

\begin{equation}
\psi_{eq}^{2} = - \frac{3b_{SWS}}{4c_{SWS}}.
\label{eq:31}
\end{equation}

The nonzero-state spinodal occurs at

\begin{equation}
a_{sp} = \frac{b_{SWS}^{2}}{4c_{SWS}}.
\label{eq:32}
\end{equation}

These equations translate the within-N3 SWS hypothesis into empirically
testable signatures. If consolidated SWS is mixed-transition-like or
near a tricritical limit, then EEG-derived \emph{$\phi$} should show a
continuous precursor through N2/N3, altered derivative structure near
the boundary, and a sharper endpoint transition or bimodality in
slow-oscillation dominance. If the entire descent remains smooth and
unimodal with no derivative change, no bimodality, and no distinct
spatial nucleation pattern, the mixed-transition hypothesis should be
weakened or rejected. Recovery of ideal mean-field exponents is not
required for support and should be treated as a secondary analysis
because PSG data are finite, noisy, and nonstationary.

This hypothesis may be relevant for glymphatic physiology, but the
connection is speculative. Rodent studies show that natural sleep and
anesthesia are associated with increased interstitial space and enhanced
convective exchange of cerebrospinal and interstitial fluid, with
enhanced clearance of amyloid beta during sleep \cite{iliff2012,xie2013,hablitz2019}. Human work has shown coupling among
slow neural activity, hemodynamics, and cerebrospinal fluid oscillations
during sleep \cite{fultz2019}, and recent model-supported human evidence
from a randomized crossover study is consistent with sleep-active
clearance of amyloid beta and tau to plasma, with relationships to NREM
duration and EEG delta power \cite{dagum2026}. These findings do not prove
that consolidated SWS is a first-order transition. They support only the
biological plausibility of coupling between deep NREM neural dynamics
and fluid-transport physiology. This literature is not unanimous: some
rodent work instead reports reduced brain clearance during sleep and
anesthesia \cite{miao2024}, and other work attributes NREM glymphatic
inflow to norepinephrine-driven vasomotion \cite{hauglund2025}, so the
effect remains measurement- and tracer-dependent and unsettled.

The proposed discontinuity should therefore be sought in measurable
order parameters or coupled physiological variables---slow-oscillation
power, Lempel-Ziv complexity, spatial coherence, CSF-flow surrogates,
heart-rate variability, noradrenergic-linked vasomotion, or parenchymal
transport markers---rather than in literal thermodynamic latent heat.
The appropriate physical analogy is a macroscopic discontinuity in
density, transport, or collective state organization, not an assertion
that the brain releases measurable latent heat in the thermodynamic
sense.

The stabilizing sixth-order term can be interpreted biologically as soft
high-amplitude stabilization of synchronization rather than a hard ceiling. Neurons cannot synchronize without limit
because of refractory periods, ionic constraints, metabolic supply,
synaptic fatigue, and homeostatic downscaling \cite{tononi2006}.
These constraints prevent the effective potential from becoming
unbounded at high \emph{$\phi$}.

\subsection{NREM-to-REM: first-order-like desynchronizing transition}\label{sec:nrem-to-rem-first-order-like-desynchronizing-transition}

One empirical caution frames this boundary. Human REM episodes
frequently emerge from N2 rather than from consolidated N3, and the
proportion shifts systematically across the night. Any
transition-centered test of an NREM-to-REM discontinuity must therefore
be stratified by the source stage, because pooling N2-to-REM and
N3-to-REM transitions can create or mask apparent bimodality and
hysteresis that reflect the mixture rather than the boundary itself.

The transition from NREM, especially SWS, to REM has theoretical support
as a discontinuous cortical transition in mean-field modeling.
Steyn-Ross and colleagues modeled SWS-to-REM as a first-order phase
transition from a low-firing coherent cortical state to a high-firing
desynchronized state driven by neuromodulatory changes, especially
ultradian cycling of acetylcholine and changes in synaptic efficiency
and resting voltage \cite{steynross2005}. REM also depends on
reciprocal inhibitory circuitry in the brainstem and hypothalamus that
can produce state switching \cite{lu2006,scammell2017}.

In the present notation, REM is modeled as a transition from high
\emph{$\phi$} to low \emph{$\phi$}, driven by a REM-biasing field. A minimal
potential is

\begin{widetext}
\begin{equation}
V_{NREM \rightarrow REM}(\psi) = \frac{a_{R}(ACh,NE)}{2}\psi^{2} + \frac{1}{4}\psi^{4} - h_{R}(ACh,NE)\psi,\quad\quad a_{R} < 0\ \text{(quartic cusp)}.
\label{eq:33}
\end{equation}
\end{widetext}

The REM transition is first-order-like, in the sense of the quartic cusp
of Sec.~\ref{sec:distinguishing-dynamical-mechanisms}, if two basins coexist: a high-\emph{$\phi$} synchronized NREM
basin and a low-\emph{$\phi$} desynchronized REM basin. The transition should
show a rapid fall in slow-wave synchronization, a rise in
mixed-frequency activity, loss of muscle tone, and a sharp change in
autonomic and sensory gating variables. Because REM is multidimensional,
the first-order prediction is strengthened only if NREM-to-REM and
REM-to-NREM paths occur at different thresholds after EEG, EOG, EMG,
autonomic state, and cycle position are modeled jointly.

\subsection{REM-to-wake: arousal as escape from a desynchronized sleep attractor}\label{sec:rem-to-wake-arousal-as-escape-from-a-desynchronized-sleep-attractor}

REM and wake share desynchronized cortical EEG features, yet they differ
strongly in neuromodulatory tone, muscle atonia, sensory gating,
autonomic patterning, responsiveness, and conscious access. REM-to-wake
transitions are therefore modeled as switches between distinct
desynchronized attractor families. The scalar cortical-ordering variable
is not sufficient for this boundary. A multidimensional order parameter
is more appropriate:

\begin{equation}
\mathbf{q}(t) = \left( \phi(t),\rho(t),\mu(t),\gamma(t) \right),
\label{eq:34}
\end{equation}

where \emph{$\rho$} represents sensory coupling, \emph{$\mu$} represents muscle
tone or motor access, and \emph{$\gamma$} represents aminergic arousal tone. A
local quadratic approximation near a REM basin is

\begin{equation}
V_{REM}\left( \mathbf{q} \right) = \frac{1}{2}\left( \mathbf{q} - \mathbf{q}_{R} \right)^{T}\mathbf{H}_{R}\left( \mathbf{q} - \mathbf{q}_{R} \right) - \mathbf{h}_{W} \cdot \mathbf{q},
\label{eq:35}
\end{equation}

where $\mathbf{H}_R$ is the local curvature matrix and $\mathbf{h}_W$ is the wake-biasing
field. Equation~(\ref{eq:35}) is a local, linear-response approximation of the
geometry \emph{within} the REM basin: for a stable basin $\mathbf{H}_R$ is
positive definite, so the potential has a single minimum at
$\mathbf{q}_{*} = \mathbf{q}_R + \mathbf{H}_R^{-1}\mathbf{h}_W$, with no
barrier or second attractor. It therefore describes how a bias field
displaces the REM minimum, not the arousal transition itself. Representing
REM-to-wake escape requires a two-basin (nonlinear) structure in the
multidimensional state vector that this local quadratic form does not
capture. The transition may then be abrupt even if cortical EEG
desynchronization changes little, because the discontinuity lies in that
multidimensional structure rather than in \emph{$\phi$} alone.

\begin{table*}[t]
\caption{Hypothesized transition taxonomy, evidential status, and key empirical tests.}
\label{tab:taxonomy}
\footnotesize
\setlength{\tabcolsep}{4pt}
\noindent\begin{tabular}{@{}p{0.20\textwidth}p{0.25\textwidth}p{0.25\textwidth}p{0.20\textwidth}@{}}
\hline\hline
\textbf{Transition} & \textbf{Hypothesized class} & \textbf{Primary observables} & \textbf{Evidential status and strongest test} \\
\hline
Wake to N1 & Local fold (saddle-node); nested question of a global bistable cusp & alpha-theta balance; inverse complexity; EOG/EMG/arousal coupling & Fold externally supported (Li \textit{et al.}); cusp embedding under-tested. Test tipping and critical slowing for the fold, and matched-control bimodality, hysteresis, and flickering for the cusp \\
N1 to N2 & Continuous thalamocortical ordering or rounded crossover & spindle density; sigma power; K-complex probability; entropy & Plausible but scoring-confounded; test continuous graphoelement probability rather than first scored occurrence \\
N2 to N3 & Continuous slow-wave recruitment & slow-wave power; inverse complexity; phase synchrony; spatial correlation length & Strongest candidate; requires controls for AASM slow-wave thresholds and preferably multi-channel/high-density EEG \\
Within N3/consolidated SWS & Tricritical limiting onset (continuous); a mixed continuous-plus-switch subregime is a separate, speculative possibility & very-slow (0.1--0.5 Hz) dominance; low LZC; spatial coherence; physiological transport surrogates & Speculative; first establish a distinct within-N3 substate and then test derivative change, bimodality, or nucleation \\
NREM to REM & First-order-like desynchronizing transition & slow-wave collapse; mixed-frequency activity; EOG/EMG; autonomic state & Plausible; must be tested as a multidimensional EEG/EOG/EMG transition, not cortical \emph{$\phi$} alone \\
REM to wake & Multidimensional arousal switch & sensory coupling; muscle tone; aminergic/arousal proxies; responsiveness & Plausible but outside scalar model; requires state vector including motor and sensory access \\
\hline\hline
\end{tabular}
\end{table*}
\section{Illustrative numerical examples: one model family can generate the proposed signatures}\label{sec:illustrative-numerical-examples-one-model-can-generate-the-proposed-signatures}

The taxonomy of Sec.~\ref{sec:proposed-taxonomy-of-sleep-stage-transitions}
assigns different transition hypotheses to different stage boundaries. A
natural objection is that this assignment is merely descriptive, attaching
distinct labels to distinct boundaries. This section shows a narrower point:
one time-dependent Ginzburg--Landau (Langevin) equation family,
Eq.~(\ref{eq:8}), can generate stylized versions of the four proposed
signature classes as its control parameters move through the regimes
identified above. These simulations are illustrative examples and an
internal-consistency check. They are not a fit to empirical sleep data and do
not validate the transition taxonomy.

\begin{figure*}[t]
\centering
\includegraphics[width=\textwidth]{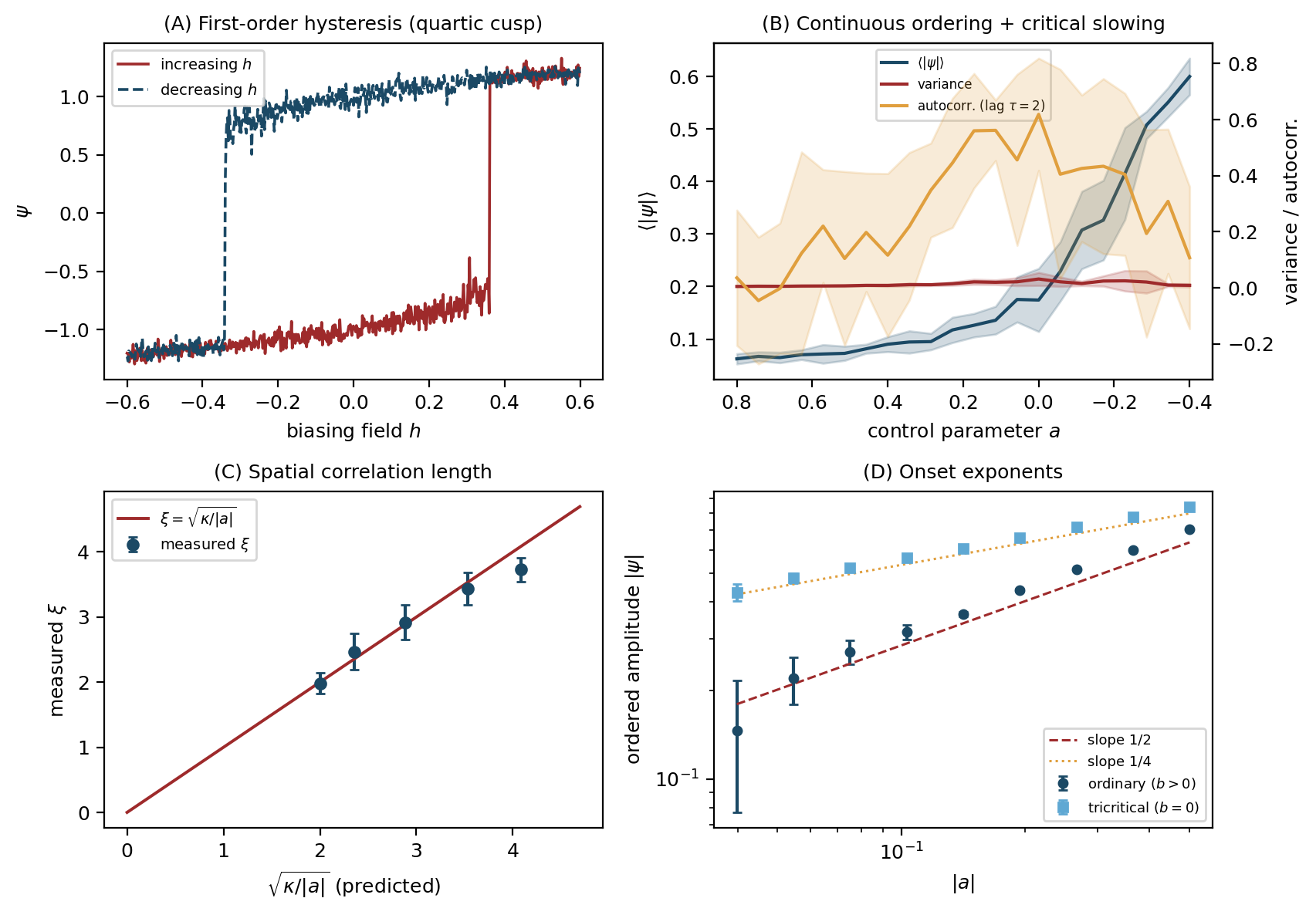}
\caption{Illustrative numerical signatures generated by the time-dependent Ginzburg--Landau family: (a) first-order-like hysteresis in a swept quartic cusp; (b) continuous ordering with critical-like slowing, in which the autocorrelation at lag $\ell$ = 2 time units is the primary resolved signature and the variance trend is weaker; (c) growth of spatial correlation length, compared with the analytic relation $\xi=\sqrt{\kappa/|a|}$; and (d) ordinary and tricritical onset exponents, with the mean-field slopes 1/2 and 1/4 drawn as reference lines. The measured onset exponents were 0.58 and 0.26. The simulations are internal-consistency examples, not fits to sleep data.}
\label{fig:sig}
\end{figure*}

The dynamics were integrated with the Euler-Maruyama scheme,

\begin{equation}
\psi_{t + \Delta t} = \psi_{t} - \Gamma\,\Delta t\left\lbrack a\psi_{t} + b\psi_{t}^{3} + c\psi_{t}^{5} - h \right\rbrack + \sqrt{2D\,\Delta t}\mspace{6mu}\xi_{t},
\label{eq:36}
\end{equation}

where the stochastic increment is standard Gaussian noise scaled by the
noise intensity; temporal signatures used the zero-dimensional
reduction; the spatial signature added the diffusive Ginzburg term on a
one-dimensional cortical line with periodic boundaries. Results are
summarized in Fig.~\ref{fig:sig}.

First-order-like bistability and hysteresis (Fig.~\ref{fig:sig}A). For the quartic
cusp \(V(\psi) = \frac{a}{2}\psi^{2} + \frac{1}{4}\psi^{4} - h\psi\)
with \emph{a} \textless{} 0, the potential has two separated minima.
Sweeping the biasing field quasi-statically upward and then downward
drives a discontinuous jump at a forward threshold that differs from the
reverse threshold. The order parameter therefore traces a hysteresis
loop: the state occupied by the system depends on the direction of
approach, not only on the instantaneous control value. This is the
stylized behavior of the first-order cusp---the nested bistable-cusp
hypothesis at wake-to-N1 (as distinct from the fold, whose signature is
critical slowing rather than a hysteresis loop) and the primary model at
the NREM-to-REM boundary---and it illustrates how an abruptly scored
transition can coexist with a gradual neurochemical run-up.

Continuous ordering and critical-like slowing (Fig.~\ref{fig:sig}B). With \emph{b} \textgreater{} 0 and \emph{h} approximately zero, the same equation produces a
continuous-like ordering crossover. As the control parameter decreases
through zero, the stationary order parameter grows smoothly from zero.
Approaching the boundary from the disordered side, fluctuation variance
and autocorrelation at lag $\ell$ = 2 rise as the local curvature
decreases. These are the stylized signatures proposed for N1-to-N2 and
N2-to-N3, but in real PSG they must be separated from filtering effects,
nonstationary drift, and scoring thresholds.

Growth of the spatial correlation length (Fig.~\ref{fig:sig}C). Adding the
Ginzburg gradient term makes the transition spatially extended. The
spatial autocorrelation of fluctuations grows as the local stiffness
decreases, tracking the predicted scaling of correlation length with
\emph{$\kappa$} and a. This illustrates the local-to-global recruitment
hypothesized for the N2-to-N3 boundary, in which slow waves begin
locally and enlist larger cortical territories. This signature requires
high-density rather than single-channel EEG.

Tricritical limiting onset (Fig.~\ref{fig:sig}D). At the tricritical limit ($b \to 0$, $c>0$), the ordered amplitude grows continuously with the steeper
mean-field exponent $1/4$, a sharper onset than the ordinary continuous case.
A continuous precursor followed by a discrete endpoint would instead require
a control path entering the $b<0$ sector or an additional switching degree of
freedom; that mixed continuous-plus-switch regime is not simulated in
Fig.~\ref{fig:sig}D. This illustrates the limiting mathematical behavior
motivating the consolidated-SWS hypothesis. It should not be
read as evidence that human N3 is tuned to a tricritical point;
empirical support would require a distinguishable within-N3 substate
with derivative change, bimodality, path dependence, or local-to-global
nucleation.

Two points follow. First, the four signature classes are not independent
assumptions; they are different limits of one phenomenological family,
selected by the sign of \emph{b}, the effective field \emph{h}, the
presence of the gradient term, and the direction of control sweep.
Second, the diagnostics used here---hysteresis-loop width, fluctuation
variance, autocorrelation at lag $\ell$ = 2, correlation length, and
onset steepness---correspond to empirical estimators proposed in
Secs.~\ref{sec:operational-eeg-metrics-and-falsifiable-predictions} and \ref{sec:proposed-methods-for-empirical-validation}. The model therefore links transition classes to
analysis targets, but the existence of these signatures in simulated
dynamics does not establish their presence in real sleep.

Table~\ref{tab:params} gives the integration parameters for each panel of Fig.~\ref{fig:sig},
generated directly from the released code so the record cannot drift.
Common to all panels: integration step $\Delta$\emph{t} = 0.01, relaxation rate
\emph{$\Gamma$} = 1, additive white noise, and fixed seeds.

\begin{table*}[t]
\caption{Numerical parameters for the illustrative simulations in Fig.~\ref{fig:sig}.}
\label{tab:params}
\footnotesize
\setlength{\tabcolsep}{4pt}
\noindent\begin{tabular}{@{}p{0.20\textwidth}p{0.25\textwidth}p{0.25\textwidth}p{0.20\textwidth}@{}}
\hline\hline
\textbf{Panel} & \textbf{Potential/dynamics} & \textbf{Key parameters} & \textbf{Realizations} \\
\hline
A---hysteresis & quartic cusp, \(\dot{\psi} = - \left( \psi^{3} + a\psi - h \right)\) & \emph{a} = -1.0, \emph{D} = 5$\times$10\textsuperscript{-3}, quasi-static sweep d\emph{h}/d\emph{t} = 0.0002 per direction, \emph{h} $\in$ {[}-0.6, 0.6{]} and back (\textasciitilde599k steps per direction) & one representative sweep \\
B---continuous/critical slowing & pitchfork, \(\dot{\psi} = - \left( a\psi + b\psi^{3} \right)\) & \emph{b} = 1.0, \emph{D} = 5$\times$10\textsuperscript{-3}, \emph{a} from 0.8 to -0.4, 3000 steps after 1500-step burn-in, autocorrelation at lag $\ell$ = 2 & 12 \\
C---correlation length & linearized, \(\dot{\psi} = - \left( a\psi - \kappa\nabla^{2}\psi \right)\) & \emph{$\kappa$} = 1, \emph{N} = 256 (periodic, finite-difference Laplacian), \emph{D} = 5$\times$10\textsuperscript{-3}, \emph{a} $\in$ \{0.25, 0.18, 0.12, 0.08, 0.06\}, burn-in $8\tau_{\mathrm{rel}}$ ($\tau_{\mathrm{rel}}=1/a$), 20 samples, \emph{$\xi$} from exp fit over 0.1 \textless{} \emph{C} \textless{} 0.85, \emph{r} \textless{} \emph{N}/4 & 5 \\
D---onset exponents & \(\dot{\psi} = - \left( - |a|\psi + b\psi^{3} + c\psi^{5} \right)\) & \textbar{}\emph{a}\textbar{} $\in$ {[}0.04, 0.50{]}, \emph{D} = 2$\times$10\textsuperscript{-4}, 3000 steps after 2000-step burn-in; ordinary \emph{b} = 1, \emph{c} = 0 and tricritical \emph{b} = 0, \emph{c} = 1 & 5 \\
\hline\hline
\end{tabular}
\end{table*}

The measured onset exponents were 0.58 (ordinary; mean-field value 0.50)
and 0.26 (tricritical; 0.25), and the measured correlation length
tracked the predicted \(\xi = \sqrt{\kappa/|a|}\) to within about ten
percent (1.98, 2.47, 2.92, 3.44, 3.73 against 2, 2.36, 2.89, 3.54,
4.08).

\begin{figure*}[t]
\centering
\includegraphics[width=\textwidth]{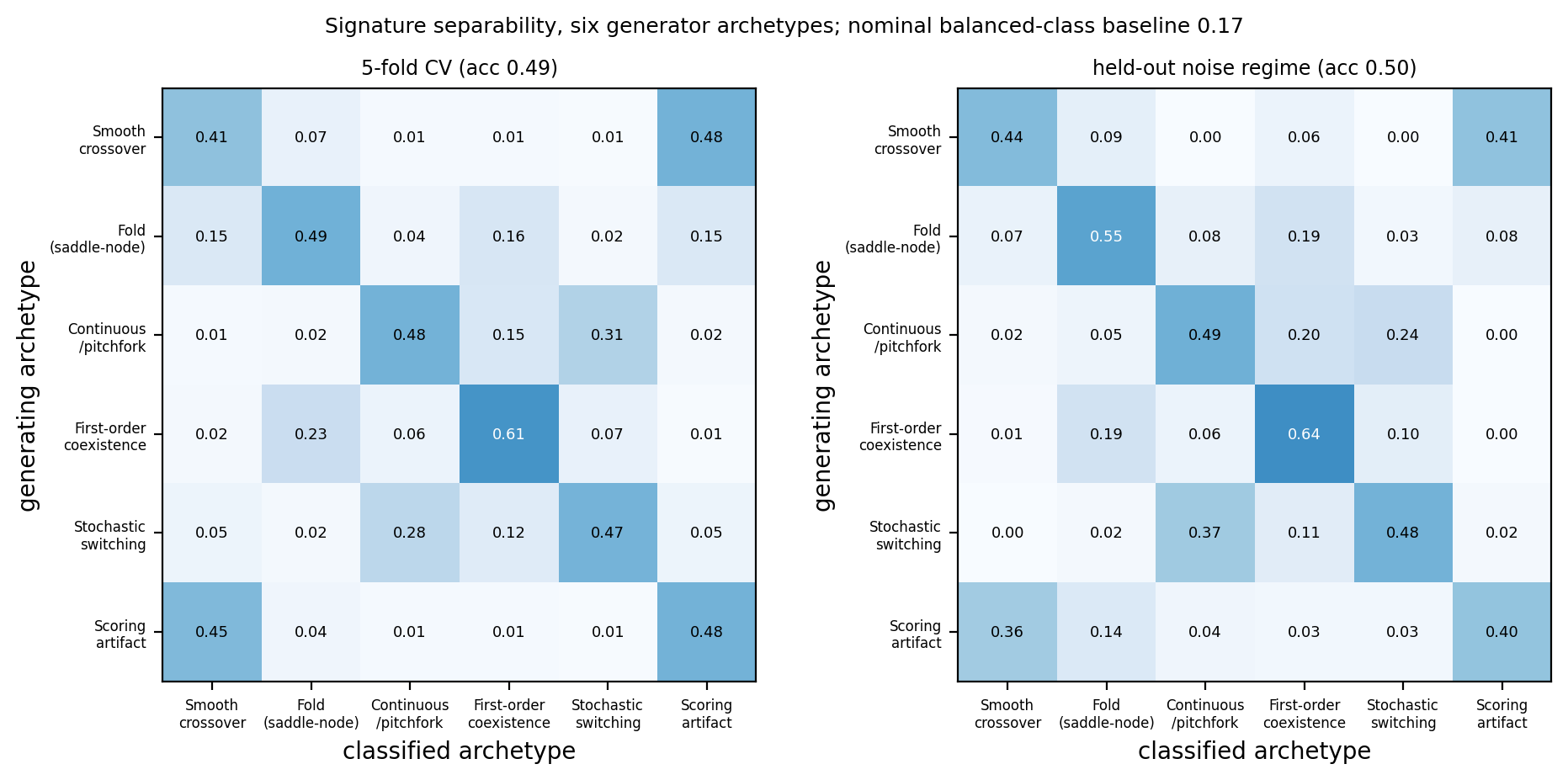}
\caption{Signature-separability confusion matrices for six candidate archetypes. Left: five-fold cross-validation (accuracy 0.49). Right: a held-out test in which the noise intensity of every archetype was increased by forty percent (accuracy 0.50). Rows are the true generating archetype; columns are the archetype recovered by a multinomial classifier with all preprocessing fitted inside each fold, archetype parameters varied across realizations, overlapping noise ranges, and the scoring-artifact archetype sharing the smooth crossover's latent dynamics. Values are row-normalized recall; chance is 0.17. The least separable pair is the smooth crossover and the scoring artifact, which share latent dynamics and differ only in scoring.}
\label{fig:recovery}
\end{figure*}

Beyond reproducing the qualitative signatures, the same dynamical
framework was used to test whether the analysis pipeline can recover the
generating archetype from synthetically degraded, non-calibrated observations. Trajectories were
generated under six idealized generator archetypes: a smooth crossover, a deterministic fold
(saddle-node), a continuous pitchfork with critical slowing, first-order
coexistence with a swept field, metastable stochastic switching at fixed
control, and a scoring-induced discontinuity. Two features of the design
keep the test from being artificially easy. First, the scoring-artifact
class is generated from exactly the same latent-trajectory distribution
as the smooth crossover, differing only in the scoring operation, so any
separation between them must come from scoring rather than from a
difference in latent shape. Second, the noise ranges of the classes
overlap, so that noise intensity is no longer determinative of the class
label, although stochastic switching still occupies the upper part of
the shared range (\emph{D} = 0.025--0.050 versus 0.010--0.040 for the
other generators). Every class additionally varies its coefficients,
ramp rate, measurement loading, and filter window from realization to
realization. Each trajectory was passed through the measurement model of
Sec.~\ref{sec:measurement-model-for-the-latent-order-parameter}, low-pass filtered, converted to categorical labels with
jittered thresholds (a coarse, dimensionless analog of epoch scoring),
and block-averaged. From the resulting continuous and categorical series,
a signature vector was computed (variance rise, change in
autocorrelation at a fixed lag, a bimodality index, dwell-time
dispersion, trend strength, and the sizes of the continuous and
categorical jumps). Each archetype contributed 150 independently drawn trajectories of 1500
steps (900 balanced trajectories in training; a separate 100 per archetype
in the shifted-noise held-out set), and each trajectory used independent parameter and
noise draws, so no group of related trajectories could leak across folds. Classification used $L_2$-penalized multinomial logistic
regression (lbfgs solver, inverse-regularization $C=1$) with standardization
fitted inside each fold of a five-fold stratified split; accuracy is reported
as plain and balanced accuracy, which coincide here because the classes are
balanced. Under
five-fold cross-validation, the pipeline recovered the correct archetype
with an accuracy of 0.49 against a nominal balanced-class baseline of 0.17
(Fig.~\ref{fig:recovery}), roughly three times the baseline rate. Per-class recall ranged from 0.61 for
first-order coexistence down to 0.41 for the smooth crossover, which is
confused mainly with the scoring artifact that shares its latent
dynamics. That confusion is the honest and expected outcome: a smooth
crossover and a scoring-induced jump on the same underlying signal are
 genuinely
hard to separate, which is precisely the circularity the validation
program is designed to guard against. To probe robustness, the
classifier was then applied to a held-out set in which the noise
intensity of every class, including stochastic switching, was increased
by forty percent; accuracy was essentially unchanged at 0.50, and the
full held-out confusion matrix is shown alongside the cross-validated
one in Fig.~\ref{fig:recovery}. The result is deliberately bounded: these six archetypes are distinguishable above the balanced-class baseline
from short, noisily observed, coarsely scored windows, this holds under a shift of noise
regime, and the least separable pair is the smooth crossover and its
scoring artifact. These six are idealized generator regimes, or
signature archetypes, rather than mutually exclusive biological
mechanisms: a single physiological window may combine several---for
example, a bistable cusp that escapes by noise before reaching its
fold---and the scoring-induced discontinuity is an observation-layer
operation, not a neural mechanism at all. The experiment therefore
measures how separable these archetypal signatures are under realistic
degradation, not that a unique mechanism can be identified in real
sleep.

To confirm that these values are stable rather than artifacts of a
single random seed, the reported quantities were recomputed across
independent master seeds (six for the correlation-length and onset-exponent
panels, five for the recovery experiment; the hysteresis panel is a single
representative sweep) and are summarized as mean $\pm$ standard deviation.
The measured onset exponents were 0.56 $\pm$ 0.05 (ordinary) and 0.25 $\pm$ 0.01
(tricritical), bracketing the mean-field values 1/2 and 1/4; the
measured correlation length tracked $\sqrt{\kappa/|a|}$ with a
seed-to-seed standard deviation below 0.24 across all five conditions;
and the signature-separability accuracy was 0.49 $\pm$ 0.005 under cross-validation
and 0.49 $\pm$ 0.01 on the shifted-noise held-out set (the single-seed value quoted above and shown in Fig.~\ref{fig:recovery} is 0.50). The reported figures
therefore reflect stable summaries rather than single-seed fluctuations.
Consistent with the framework's stated scope, these simulations remain
illustrative and non-inferential: they establish that the proposed
observables behave as claimed on synthetic data, not that any specific
numerical value characterizes empirical sleep.

\section{Motivating transition-level EEG observations}\label{sec:motivating-transition-level-eeg-observations}

A related transition-level EEG analysis by Passaro and Poltorak~\cite{passaro2025}
examined the Stanford Technology Analytics and Genomics in Sleep
(STAGES) cohort using high-resolution spectral analysis across
wake-to-N1, N1-to-N2, N2-to-N3, and N2-to-REM windows. The analysis
extracted five-minute windows on each side of each transition, applied
multitaper time-frequency analysis from 0.1 to 50 Hz, and tested
frequency-specific trends and correlations with N3 percentage, REM
percentage, total sleep time, and age. Because this work is currently a
preprint and because spectral power alone cannot identify transition
order, it is used here only as motivating evidence for the feasibility
of transition-centered EEG analysis.

Several observations are directionally consistent with the present
framework. Wake-to-N1 and N1-to-N2 were accompanied by reductions in
beta and gamma power with increases in lower-frequency activity,
indicating movement away from a high-dimensional activated state.
N2-to-N3 showed increased delta power and reductions in alpha, beta, and
gamma power, consistent with increasing synchronization. N2-to-REM
showed the opposite direction: reduced delta and sigma power with
increased beta and gamma activity, consistent with a desynchronizing
transition. The same analysis reported that elevated sigma, beta, and
gamma power in lighter stages was associated with reduced N3 percentage,
while pre-sleep delta power was associated with REM percentage and total
sleep duration.

These findings should be treated as preliminary alignment rather than
confirmation. A rigorous test requires additional order parameters,
independent transition-time estimation, microstate analysis, hysteresis
estimates, scaling analysis, null models for scoring artifacts, and
model comparison. The value of the STAGES-style approach is that it
demonstrates the feasibility of transition-centered analysis on large
PSG cohorts and identifies transition windows where the Landau--Ginzburg
predictions can be evaluated.

\section{Operational EEG metrics and falsifiable predictions}\label{sec:operational-eeg-metrics-and-falsifiable-predictions}

\subsection{Candidate order parameters}\label{sec:candidate-order-parameters}

The candidate order parameters, their concrete estimators, and their
principal caveats are summarized in Table~\ref{tab:obs}. They are
organized into three tiers: primary indicators of cortical ordering
(slow-oscillation dominance, inverse complexity, and, where spatial data
permit, synchrony and correlation length), secondary diagnostics
(aperiodic spectral slope and critical-slowing indicators), and
exploratory measures (branching dynamics), together with the
multidimensional arousal and motor-access variables required at the
REM and wake boundaries.

\begin{table*}[t]
\caption{Primary, secondary, and exploratory EEG/PSG observables for estimating transition dynamics.}
\label{tab:obs}
\footnotesize
\setlength{\tabcolsep}{4pt}
\noindent\begin{tabular}{@{}p{0.20\textwidth}p{0.25\textwidth}p{0.25\textwidth}p{0.20\textwidth}@{}}
\hline\hline
\textbf{Tier} & \textbf{Construct} & \textbf{Estimator} & \textbf{Use and caveat} \\
\hline
Primary for NREM \emph{$\phi$} & slow-oscillation dominance & 0.5--4 Hz delta power (broad); slow oscillations ($\approx$0.5--1 Hz) and very-slow activity (0.1--1 Hz) reported separately & Core indicator of NREM deepening; very low-frequency estimates require longer windows \\
Primary for NREM \emph{$\phi$} & inverse complexity & 1 - normalized Lempel-Ziv complexity; 1 - normalized permutation entropy, or $-z$-scored multiscale entropy (negative loading) & Useful for NREM ordering; window length and binarization must be prespecified \\
Primary where spatial data exist & synchrony & coherence, wPLI, phase-locking measures & Synchrony indices; not direct estimators of a cortical correlation length \\
Primary where spatial data exist & correlation length & decay of source-space covariance or phase correlation with cortical geodesic distance; fitted spatial-kernel scale & Needed for Ginzburg predictions; sparse PSG is insufficient for a true correlation length \\
Secondary diagnostic & aperiodic exponent and offset (putative E--I-related diagnostic) & spectral exponent and offset using specparam/FOOOF or related methods & Not a direct measurement of synaptic E/I balance; controls broadband slope changes that masquerade as oscillatory power \\
Secondary diagnostic & critical-slowing indicators & variance, lag-1 autocorrelation, recovery time after microarousals & Must control filtering, nonstationarity, and sleep-cycle drift \\
Exploratory & branching dynamics & avalanche-size distribution; branching ratio sigma; deviation from criticality & Best for dense/intracranial data; fragile in sparse scalp PSG \\
Multidimensional REM/wake & arousal and motor-access variables & EOG, chin EMG, autonomic signals, responsiveness when available & Required for REM-wake transitions because cortical \emph{$\phi$} alone is insufficient \\
\hline\hline
\end{tabular}
\end{table*}
\subsection{Predictions distinguishing transition classes}\label{sec:predictions-distinguishing-transition-classes}

Prediction 1a (local fold): Wake-to-N1 should show a reproducible tipping
point with critical slowing---rising variance and autocorrelation---before
loss of wake stability, more clearly than at N1-to-N2 or N2-to-N3.

Prediction 1b (conditional cusp): If that fold is embedded in a bistable
cusp, matched-control wake-entry and wake-recovery paths should differ,
with bimodality, metastability, or path dependence, so that the same
subject crosses into sleep at a different order-parameter threshold than
the one at which microarousal restores wakefulness. Absent the cusp
embedding, the local fold alone does not predict hysteresis. Flickering switches consistent with this
prediction have already been reported at small scale~\cite{hu2026}.

Prediction 2: N1-to-N2 and N2-to-N3 should show relatively smooth
changes in complexity, slow-wave power, and spatial synchrony. Evidence
for a true discontinuity at these boundaries would weaken the proposed
taxonomy.

Prediction 3: If consolidated SWS approaches a tricritical-like limit,
slow-oscillation dominance and inverse complexity should show a continuous but
steeper onset. A genuinely mixed continuous-plus-switch subregime would
additionally show a sharper derivative change, a local-to-global nucleation
pattern, a distinct entry/exit threshold, or a bimodal distribution near the
onset of sustained global slow oscillations. Exact tricritical scaling is a secondary and
unlikely-to-be-clean endpoint in standard PSG.

Prediction 4: NREM-to-REM transitions should show a larger
multidimensional discontinuity than the N2-to-N3 boundary, including
rapid changes in slow-wave synchronization, mixed-frequency activity,
muscle tone, eye movements, and autonomic/arousal variables. Entry and
exit thresholds should differ across the ultradian cycle after
controlling for sleep pressure and clock time.

Prediction 5: Age, insomnia, and neurodegenerative vulnerability may be
expressible as changes in effective model coefficients rather than
simply as reductions in stage duration. For example, insomnia may deepen
the wake basin or shift the saddle-node-like threshold; aging may reduce
the attainable maximum of \emph{$\phi$}, reduce the spatial coupling \emph{$\kappa$}
so that the correlation length shortens and global slow-wave recruitment
becomes harder, or destabilize a high-coherence SWS subregime. These are
hypotheses for future validation, not established clinical biomarkers.

\begin{figure*}[t]
\centering
\includegraphics[width=\textwidth]{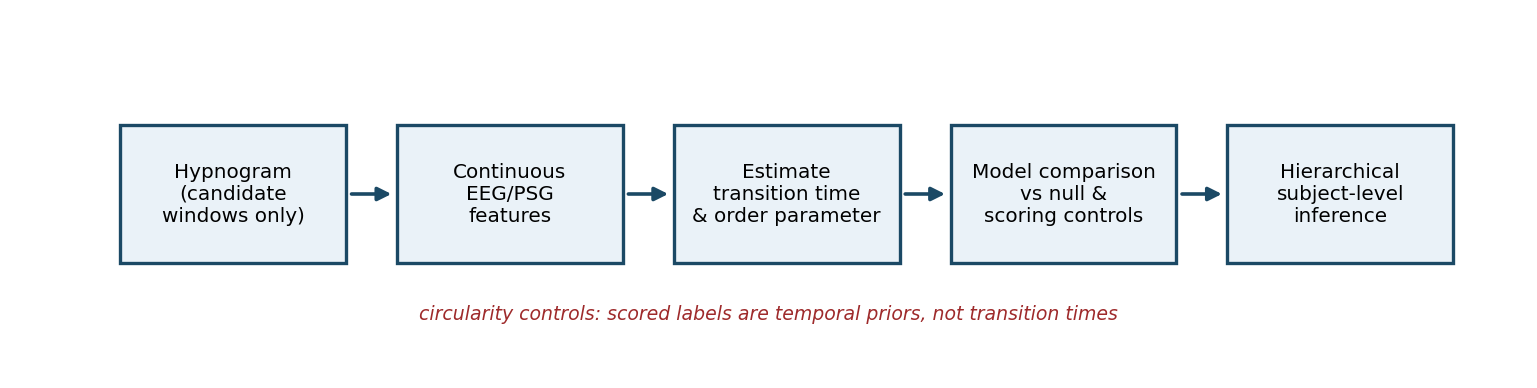}
\caption{Proposed validation pipeline with circularity controls. Standard hypnograms identify candidate transition windows only as temporal priors. Transition times and transition classes are then inferred from continuous EEG/PSG features, tested against null models and scoring-artifact controls, and aggregated with hierarchical subject-level statistics. The localization/testing split occurs inside the transition-estimation block, using disjoint features or cross-fitting to avoid selection bias.}
\label{fig:pipeline}
\end{figure*}

\begin{table*}[t]
\caption{Falsifiable statistical tests for transition order across the canonical boundaries, with key controls and data requirements.}
\label{tab:tests}
\footnotesize
\setlength{\tabcolsep}{4pt}
\noindent\begin{tabular}{@{}p{0.20\textwidth}p{0.25\textwidth}p{0.25\textwidth}p{0.20\textwidth}@{}}
\hline\hline
\textbf{Hypothesis} & \textbf{Primary test} & \textbf{Supportive/falsifying pattern} & \textbf{Main control or data requirement} \\
\hline
Wake-to-N1: local fold, nested bistable-cusp question & Fold: tipping-point and critical-slowing tests. Cusp: mixture/change-point models; entry-exit hysteresis; dwell-time distributions & Fold support: reproducible tipping point with rising variance and autocorrelation. Cusp support: two-state distribution, distinct thresholds, flickering. Falsify cusp: unimodal smooth trajectory with no path dependence. & Control eye closure, alpha dropout, scoring delay, and N1 instability; include EOG/EMG where possible. \\
N2-to-N3 is continuous-like & smooth vs switching model comparison; variance/autocorrelation; spatial coherence & Support: smooth order-parameter change with scale-dependent fluctuation. Falsify: large jump and stable bimodality independent of scoring. & Control AASM slow-wave threshold; use continuous slow-wave density rather than labels alone. \\
Within-N3 SWS is mixed-transition-like & derivative change; bimodality in SO dominance; local nucleation maps; entry/exit thresholds & Support: continuous precursor plus sharper endpoint or nucleation. Falsify: no distinguishable SWS subregime inside N3. & Requires high-density EEG or R\&K Stage 3/4 as an approximation; glymphatic claims need physiological measures. \\
NREM-to-REM is first-order-like & jump magnitude; hysteresis across cycle phase; rapid collapse of SO synchrony plus EOG/EMG change & Support: abrupt multidimensional switch and separate REM basin. Falsify: gradual reversible trajectory with no discontinuity. & Model EEG, EOG, EMG, autonomic variables, sleep pressure, and circadian/cycle phase jointly. \\
N1-to-N2 is continuous-like & smooth vs.\ switching model comparison; spindle/SO onset; variance and autocorrelation & Support: smooth spindle/SO growth with no jump. Falsify: discrete jump or stable bimodality independent of scoring. & Use continuous spindle/SO density rather than labels; control spindle-detector thresholds. \\\\
REM-to-wake: multidimensional arousal transition (trigger may be deterministic stability loss or noise-assisted escape) & compare recovery/slowing before arousal, fixed-threshold vs.\ noise-dependent timing, and a multidimensional switching model; dwell-time/escape-rate statistics & Support (escape): escape statistics that scale with arousal drive. Support (deterministic): critical slowing before a fixed threshold. & Model EEG with EOG/EMG and autonomic arousal; separate spontaneous from stimulus-evoked awakenings. \\
\hline\hline
\end{tabular}
\end{table*}
\subsection{Confounds and identifiability}\label{sec:confounds-and-identifiability}

Each predicted signature has a mundane alternative explanation, and the
framework's credibility depends on ruling those out rather than on
observing the signature. Bimodality can arise from pooling subjects,
cycles, or nonstationary segments rather than from genuine coexistence;
the test is whether bimodality survives within-subject, within-cycle
analysis and matched control values. Apparent hysteresis can be produced
by drift in homeostatic pressure, circadian phase, or prior-state
duration rather than by a bistable potential. This is the single most
demanding inference in the framework, because forward and reverse
transitions in natural sleep occur under different, partly unobserved
homeostatic and circadian conditions: non-overlapping forward and
reverse paths therefore establish path dependence but not, by
themselves, hysteresis from a bistable potential. Licensing the stronger
inference requires measured control variables (for example, an estimate
of homeostatic pressure from prior sleep-wake history and a circadian
phase marker), an explicit matching or regression procedure that equates
those variables across the forward and reverse directions, negative
controls in which no bistability is expected, and ideally a within-night
or interventional protocol that revisits the same boundary under matched
conditions. Absent these, we report only path dependence. A rise in
variance and lag-1 autocorrelation can come from filtering, changing
spectral composition, respiratory artifact, or slow drift; the test is
whether the rise is specific to approaching the boundary and absent
in matched non-transition segments. Correlation-length growth can be
mimicked by a shared global drive or by volume conduction; the test is
whether it appears in source-reconstructed or Laplacian-referenced
signals and scales as predicted with the gradient term.

The general principle is that no single signature identifies a
mechanism; identification requires the joint pattern specified in
Sec.~\ref{sec:distinguishing-dynamical-mechanisms}. In particular, the observations must be able to separate a
smooth nonlinear trajectory, a scoring-imposed change point, a
hidden-Markov switch, passage through a saddle-node, noise-induced
escape, first-order coexistence, and a rounded continuous transition,
and the analysis should report which of these the data can and cannot
distinguish rather than asserting a single winner.

Because no equilibrium
fluctuation-dissipation relation is assumed, susceptibility cannot be
inferred directly from spontaneous variance: a large variance need not imply a
large response to perturbation. A genuine susceptibility claim requires
either an actual perturbation, for example a sensory or electrical probe
with a measured response, or a specified stochastic model that connects
the noise covariance to the curvature of the effective potential.
Spontaneous-variance and response-based quantities should not be
conflated.

\section{Proposed methods for empirical validation}\label{sec:proposed-methods-for-empirical-validation}

\subsection{Datasets}\label{sec:datasets}

The framework can be tested initially with public and semi-public PSG
datasets. Sleep-EDF Expanded provides accessible overnight EEG and
hypnograms and is useful because the original R\&K
labels preserve Stage 3 and Stage 4 separately \cite{kemp2000}. This
permits a practical first test of the proposed within-N3
consolidated-SWS boundary using Stage 3-to-Stage 4 transitions, while
acknowledging that this mapping is an operational approximation rather
than a claim that R\&K Stage 4 is identical to the proposed substate.
STAGES provides a much larger clinical PSG resource and allows
transition-level cohort analyses when access is available. High-density
EEG datasets are especially valuable for spatial correlation length,
traveling waves, and local sleep nucleation.

The unit of analysis should be the transition event rather than the
30-second epoch. Candidate transitions are identified from expert-scored
hypnograms and then reanalyzed in overlapping windows centered on the
boundary. Because scored stage boundaries are coarse and rule-based, the
statistical transition time should be estimated from EEG/PSG order
parameters independently of the scored boundary, then compared with the
scored boundary as an external reference.

\subsection{Preprocessing}\label{sec:preprocessing}

EEG should be filtered conservatively, artifact-rejected, and referenced
consistently. Eye-movement, EMG, respiratory, and cardiac channels
should be retained where possible because REM/wake transitions are
multidimensional. The analysis should avoid imposing overly aggressive
stationarity assumptions. Window duration should be metric-specific: 2--5
s windows can capture abrupt spectral shifts, alpha dropout, spindles,
and K-complexes; 10--30 s windows are more appropriate for slow-wave
dominance, entropy, and aperiodic spectral fitting; and DFA/Hurst,
avalanche analyses, and robust critical-slowing estimates require longer
segments or event ensembles. Very low-frequency estimates require longer
windows or model-based approaches.

Periodic and aperiodic components should be separated when possible,
because changes in broadband spectral slope can masquerade as changes in
oscillatory power. The specparam/FOOOF approach or related spectral
parameterization methods can quantify exponent, offset, and oscillatory
peaks \cite{donoghue2020,lendner2020}.

\subsection{Transition-centered statistics}\label{sec:transition-centered-statistics}

For each transition window, compute prespecified order-parameter
estimators $\phi_{k}(t)$ across channels and spatial scales. Fit three
families of models: continuous smooth trajectories, such as generalized
additive models or sigmoid growth; discontinuous or switching models,
such as hidden Markov models, change-point models, or mixture models;
and mixed models with a continuous precursor and discrete switch. Model
comparison should be performed within subjects and then aggregated with
hierarchical statistics. To avoid selection bias from localizing the
transition and testing its shape with the same features, the transition
time is estimated from a prespecified feature subset (or a held-out set of
channels or source regions) disjoint from the features used to test the
mechanism, or by cross-fitting that alternates the data used for
localization and testing, with uncertainty in the estimated transition time
propagated into the model comparison. Null models should shuffle or jitter scored
boundaries, match sleep-cycle phase, and control for subject-level
baseline differences so that scoring-rule artifacts are not mistaken for
transition dynamics (Fig.~\ref{fig:pipeline}).

Hysteresis can be tested by comparing forward and reverse transitions
within the same night: wake-to-N1 versus N1-to-wake, N2-to-N3 versus
N3-to-N2, and NREM-to-REM versus REM-to-NREM. A transition is
first-order-like if the forward and reverse paths in order-parameter
space do not overlap after controlling for clock time, sleep pressure,
circadian phase, and prior sleep history. Because hidden slow variables,
adaptation, delay, and non-Markovian memory can also produce non-overlapping
paths, path dependence is treated as first-order evidence only in
combination with independent evidence for two basins, metastability, or a
fitted bistable landscape. Because natural sleep
trajectories are drifting rather than experimentally clamped, hysteresis
estimates should be interpreted as path-dependence evidence rather than
as literal equilibrium hysteresis loops.

Critical slowing down can be tested by estimating lag-1 autocorrelation,
variance, recovery time after spontaneous microarousals, and DFA scaling
\cite{peng1995} in pre-transition windows. Continuous transitions
should show scale-sensitive fluctuations around the boundary.
First-order-like transitions may show flickering between basins when the
barrier is low, along with bimodality in the order parameter, which can
be assessed with unimodality tests such as the Hartigan dip test
\cite{hartigan1985}. Rising autocorrelation is not specific,
however, and must be separated from filtering, slow drift, respiratory
artifacts, and nonstationarity.

\subsection{Decision criteria for model comparison}\label{sec:decision-criteria-for-model-comparison}

To keep the model comparison confirmatory rather than exploratory, we
fix the decision criteria in advance. Competing generative models
(smooth, switching, first-order, and mixed) are compared by
cross-validated out-of-sample predictive score, with subjects, not
epochs, as the held-out unit, so that comparisons reflect generalization
across people rather than within-night overfitting; nested models are
additionally compared by an information criterion. A boundary is
assigned a mechanism only if the winning model is preferred both by
held-out predictive score (a prespecified advantage in expected log predictive
density exceeding one standard error of the paired subject-level difference)
and by a model-appropriate secondary criterion---BIC where regularity conditions
hold, or a marginal-likelihood approximation or an information criterion suited
to latent-state and mixture models, whose usual BIC approximation can be
unreliable, and if the
corresponding signature survives the confound tests of Sec.~\ref{sec:confounds-and-identifiability}.
The Benjamini-Hochberg false-discovery-rate procedure \cite{benjamini1995}
controls the family of confound and early-warning hypothesis tests across
boundaries and metrics; the model-selection step above is governed by the
predictive-score and information-criterion rule rather than by that
correction. Measurement invariance of the order-parameter model
(Sec.~\ref{sec:measurement-model-for-the-latent-order-parameter}) is tested before any pooling across subjects, cycles,
ages, or montages, and results are reported separately where invariance
fails.

\subsection{Spatial tests of the Ginzburg term}\label{sec:spatial-tests-of-the-ginzburg-term}

A measurement caveat governs all spatial tests. The correlation length
of the order-parameter field is not the same quantity as scalp EEG
coherence, and the two should not be equated. Scalp coherence is
strongly shaped by the choice of reference, by volume conduction, and by
the scalp lead field, all of which can inflate apparent long-range
synchrony independently of any true growth in cortical correlation
length. Spatial tests of the Ginzburg term should therefore use
source-reconstructed signals, surface-Laplacian or
current-source-density montages, or explicit lead-field modeling, and
should demonstrate that any measured growth in correlation length is not
an artifact of reference and conduction.

The gradient term predicts spatial structure. If NREM deepening is
continuous, correlation length should expand gradually as local slow
waves recruit larger cortical territories. If consolidated SWS emerges
by nucleation, high-density EEG should reveal local patches of high
\emph{$\phi$} that expand as traveling slow-wave fronts. These predictions
cannot be tested adequately with one or two central EEG channels.
High-density EEG, source-localized EEG/MEG, or intracranial recordings
are needed.

Spatial analyses should estimate traveling-wave direction, phase
velocity, interregional delay, and local onset time of slow
oscillations. The key test is whether the onset of consolidated SWS
resembles gradual global alignment or nucleation followed by domain
growth.

\subsection{Empirical validation on public data}\label{sec:empirical-validation-on-public-data}

The natural empirical test of the framework is a transition-centered
analysis of public polysomnography. Clean wake-to-N1, N1-to-N2, N2-to-N3,
and NREM-to-REM transitions can be selected from a corpus such as
Sleep-EDF, where a clean transition requires a prespecified minimum dwell
time in both the source and destination stages, exclusion or modeling of
intermediate epochs and rapid reversals, removal of epochs contaminated by
scored arousals, limb movements, or respiratory events, stratification by
sleep-cycle number and, for REM, by source stage, and clustering of
multiple transitions from the same participant; a window of approximately
five minutes on each side extracted; transition times estimated independently from continuous EEG
features; a small primary metric set computed, including alpha-theta
balance, slow-wave dominance, and inverse complexity or permutation
entropy, together with EOG/EMG variables where available; and smooth,
switching, and mixed models compared with subject-level random effects.
The decisive comparison is whether wake-to-N1 and NREM-to-REM show stronger
discontinuity or bimodality than N2-to-N3. Such an analysis tests
whether the proposed observables discriminate transition classes in real
sleep, and the synthetic signature-separability result of Sec.~\ref{sec:illustrative-numerical-examples-one-model-can-generate-the-proposed-signatures} establishes
that the pipeline can make that discrimination when the ground truth is
known.

\section{Speculative clinical and translational implications}\label{sec:speculative-clinical-and-translational-implications}

The translational implications remain hypothetical. The framework recasts some
sleep disorders as alterations in state-space stability and transition
dynamics. Insomnia could reflect failure to destabilize the wake attractor;
hyperarousal, anxiety, conditioned wakefulness, pain, and circadian
misalignment may deepen the wake basin or shift the control threshold required
for sleep onset. Sleep fragmentation could reflect an abnormally low barrier
between sleep and wake basins. REM instability could reflect failure to
stabilize a REM attractor or inappropriate switching between REM and wake-like
states. Age-related loss of slow-wave sleep could reflect reduced capacity to
reach or maintain high-\emph{$\phi$} NREM regimes \cite{mander2017}.

This view suggests candidate dynamical biomarkers for future testing.
Sleep latency alone may be less informative than the slope and curvature
of the approach to the sleep-onset boundary. N3 percentage may be less
informative than the ability to enter and sustain a high-coherence
slow-oscillation regime. Arousal index may be less informative than
basin depth and recovery time after perturbation. These quantities could
in principle be estimated from EEG and autonomic signals, but the
present theoretical manuscript does not establish diagnostic biomarkers.

A related possibility is to formulate closed-loop neuromodulation as a
perturbation of transition dynamics rather than continuous entrainment to a
target frequency. Such a system might estimate the current basin, identify
proximity to a transition boundary, and apply sensory or electrical
stimulation that changes effective control parameters, reduces barrier height,
or nudges the system toward a physiological NREM trajectory. For sleep
applications, any such strategy would require empirical validation and special
caution to preserve natural ultradian architecture.

\section{Limitations and boundary conditions}\label{sec:limitations-and-boundary-conditions}

The model is phenomenological. Landau--Ginzburg coefficients do not map
one-to-one onto single neurotransmitters, and the cortical order
parameter is a low-dimensional summary of a much richer state. This is a
strength for generality and a limitation for mechanistic specificity.
Any proposed mapping from adenosine, acetylcholine, norepinephrine,
thalamocortical gain, or sleep pathology to \emph{a}, \emph{b}, \emph{c}, \emph{$\kappa$}, \emph{h}, or
\emph{$\Gamma$} must be tested empirically. REM and wake transitions require a
multidimensional state vector; the scalar \emph{$\phi$} formalism is most
appropriate for NREM ordering. More fundamentally, the framework is
local by construction: it models the dynamics within individual
transition windows and does not generate the sleep cycle itself.
Ultradian rhythmicity, REM-on/REM-off oscillation, and the asymmetric
ordering of the cycle require non-gradient drift, delays, or Hopf
components that are outside a relaxational gradient model; here the slow
control parameters are supplied externally rather than produced by the
model. Extending the description to autonomous whole-night dynamics, by
coupling the order-parameter field to explicit Process S and Process C
and REM-switch dynamics or by deriving the local normal form from an
established neural-mass model, is the natural next theoretical step and
is not attempted here.

Sleep stages are scored labels with limited temporal precision. A
30-second epoch can contain mixtures of microstates, local sleep,
K-complexes, spindles, and arousals. Transition analyses must therefore
avoid circularity: the scored hypnogram should identify candidate
windows, while the transition time and transition order should be
inferred from continuous EEG and physiological measures. Any claim of
discontinuity must be shown to survive null models that jitter, shuffle,
or re-estimate scoring boundaries.

The term phase transition is used here in an operational,
statistical-physics-inspired sense of collective state reorganization
adapted to noisy nonequilibrium biological systems. It does not imply
equilibrium thermodynamics, literal temperature, or macroscopic latent
heat. Claims about first-order-like or continuous character require
operational evidence: discontinuities, hysteresis, bimodality, scaling
behavior, correlation-length growth, or independently estimated loss of
stability. Nonzero effective fields, finite-size effects, and
measurement noise can round sharp transitions into crossovers.

The proposed tricritical SWS hypothesis is the most speculative
component of the framework. It has been reframed here as a possible
mixed-transition subregime within N3, with exact tricriticality treated
as an idealized limiting case. The hypothesis is included because it
makes concrete predictions about within-N3 structure, slow-oscillation
nucleation, and possible coupling to neural-fluid physiology. It should
be abandoned or revised if high-density EEG and physiological data fail
to identify a distinct consolidated SWS subregime with mixed-transition
signatures.

The numerical examples also have limited evidential weight. They
demonstrate that a time-dependent Landau--Ginzburg family can generate
stylized signatures, but they do not derive coefficients from neural
circuitry, fit EEG data, or establish transition order in human sleep.
The relaxational dynamics are a minimal Model-A-like approximation
\cite{hohenberg1977} to slow order-parameter evolution; full
sleep dynamics also involve delays, reciprocal inhibition, oscillations,
neuromodulatory cycles, autonomic variables, and non-gradient flows.

\section{Conclusion}\label{sec:conclusion}

A hypnogram is a useful clinical projection of a richer cortical trajectory.
The underlying dynamics may include local stability loss, bistable switching,
continuous-like ordering, and mixed-transition regimes. Landau--Ginzburg
phenomenology offers a compact language for separating these possibilities and
converting them into empirical tests.

The framework links classical sleep regulation, cortical mean-field models, criticality, and transition-centered
EEG analysis while keeping their levels of description distinct. It predicts
that major state changes may be switch-like, NREM deepening may be a rounded
ordering process, and highly consolidated SWS may contain an additional mixed
regime. Only the sleep-onset fold presently has strong direct empirical
support; the remaining assignments are hypotheses. Their value will depend on
model comparison in continuous EEG/PSG data, with scoring artifacts, hidden
control drift, and measurement circularity treated explicitly.

\section*{Funding}
No funding was received specifically for this theoretical and computational
framework paper.

\section*{Acknowledgments}
The author thanks colleagues and collaborators at NeuroLight for
discussions on sleep-stage dynamics, EEG transition analysis, and
closed-loop neuromodulation.

\section*{Conflicts of interest}
Alexander Poltorak is affiliated with NeuroLight, Inc., a company
developing neuromodulation technologies related to sleep. The company
may have intellectual-property and commercial interests related to sleep
modulation. The present manuscript is theoretical and does not establish
a clinical biomarker, diagnostic product, or validated neuromodulation
protocol.

\section*{Data and code availability}
No new empirical data were generated for this theoretical manuscript. The
numerical examples use synthetic trajectories, and their procedures and
parameters are reported in Sec.~\ref{sec:illustrative-numerical-examples-one-model-can-generate-the-proposed-signatures} and Table~\ref{tab:params}. The proposed
validation analyses can be performed on public datasets such as Sleep-EDF and
on larger datasets such as STAGES where access and data-use permissions are
available. Code for the potential diagrams, dynamical simulations, and
synthetic signature-separability experiment is provided as supplementary
material with this submission and will be deposited in a version-tagged public
repository with a citable DOI on acceptance.

\appendix
\section{Nondimensionalization and estimable parameter combinations}\label{sec:appendix}
The model is written in dimensionless form. The order parameter \emph{$\phi$}
is a normalized, dimensionless cortical-ordering coordinate; the
centered coordinate is \emph{$\psi$} = \emph{$\phi$} $-$ \emph{$\phi$}\textsubscript{0}. Space and time
are rescaled so that the gradient coefficient \emph{$\kappa$} and the
relaxation rate \emph{$\Gamma$} can be set to unity without loss of generality,
which fixes the natural length and time units of the reduced
description. The Landau coefficients then carry the remaining
dimensions: \emph{$a$} and \emph{$b$} set the curvature and the quartic character of the
potential; when present, \emph{$c$} \textgreater{} 0 provides high-amplitude stabilization; \emph{$h$} is the biasing
field; and \emph{$D$} is the noise intensity of the additive term \emph{$\eta$}. The
table records each symbol and which parameter combinations remain
estimable after rescaling; a full structural-identifiability analysis is
deferred to model fitting.

\begin{table}[!ht]
\caption{Nondimensionalization and estimable parameter combinations after rescaling.}
\label{tab:notation}
\scriptsize
\setlength{\tabcolsep}{4pt}
\renewcommand{\arraystretch}{1.15}
\noindent\begin{tabular}{@{}p{0.12\columnwidth}p{0.44\columnwidth}p{0.40\columnwidth}@{}}
\hline\hline
\textbf{Symbol} & \textbf{Meaning} & \textbf{Status after rescaling} \\
\hline
\emph{$\phi$}, \emph{$\psi$} & ordering coordinate; centered coordinate \emph{$\psi$} = \emph{$\phi$} $-$ \emph{$\phi$}\textsubscript{0} & dimensionless; \emph{$\phi$}\textsubscript{0} set by the reference state \\
\emph{$a$} & quadratic coefficient/local curvature & identifiable up to the \emph{$\phi$} and time scale \\
\emph{$b$} & quartic coefficient/transition character & enters only via scaled, dimensionless combinations; sign may be unrecoverable from short noisy trajectories \\
\emph{$c$} & sextic (saturation) coefficient & sets the order-parameter scale in the sextic sector ($c>0$); in the quartic sector $b$ plays this role; not separately identifiable from a rescaling \\
\emph{$\kappa$} & gradient (Ginzburg) coefficient & set to 1 by length rescaling \\
\emph{$\Gamma$} & relaxation rate & set to 1 by time rescaling \\
\emph{$h$} & biasing field & in the quartic-cusp sector ($a<0,\ b>0$) enters through the dimensionless field $\tilde h = h\sqrt{b}/|a|^{3/2}$ (a separate $c$-based scaling applies for $b=0$); controls rounding \\
\emph{$D$} & noise intensity & confounded with the \emph{$\phi$} scale, curvature, and sampling; not separately identifiable from short trajectories \\
\hline\hline
\end{tabular}
\end{table}

Because one nonlinear coefficient fixes the order-parameter scale ($b$ in
the quartic sector, $c$ in the sextic sector), and because
\emph{$\kappa$} and \emph{$\Gamma$} are absorbed by rescaling, only a reduced set of
dimensionless combinations---principally the sign and scaled magnitude
of \emph{b}, the cusp field $\tilde h = h\sqrt{b}/|a|^{3/2}$ (in the sector $a<0,\ b>0$), and the
scaled noise level \emph{D}---is estimable
from trajectory statistics. Setting \emph{$\kappa$} = \emph{$\Gamma$} = 1 is therefore
a choice of units, not an empirical claim: what is empirically
identifiable is not \emph{$\kappa$} or \emph{$\Gamma$} as dimensionless numbers but
the physical correlation length $\xi = \sqrt{\kappa/\mu}$ and the physical
relaxation time $\tau_{\mathrm{rec}} = 1/(\Gamma\mu)$, where
$\mu = V''(\psi_{eq})$ is the curvature (Hessian) eigenvalue at the fixed point, and the full dynamical eigenvalue is
$\lambda_{\mathrm{eig}} = -\Gamma\mu$ (so that
$\tau_{\mathrm{rec}} = 1/|\lambda_{\mathrm{eig}}|$, consistent with
Eqs.~(1)--(2)); these set the natural length and
time scales. The empirical proxies listed in Table~\ref{tab:box1} (correlation length
for \emph{$\kappa$}, recovery time for \emph{$\Gamma$}) should be read as proxies for
these dimensional scales, consistent with the rescaling used in the
simulations. Parameters that are not separately identifiable from scalp
EEG should be interpreted phenomenologically rather than as measured
physical constants.


\begin{thebibliography}{99}
\bibitem{rk1968} A.~Rechtschaffen and A.~Kales, \textit{A Manual of Standardized Terminology, Techniques and Scoring System for Sleep Stages of Human Subjects} (U.S. Government Printing Office, Washington, DC, 1968).
\bibitem{silber2007} M.~H. Silber, S.~Ancoli-Israel, M.~H. Bonnet, \textit{et al.}, J. Clin. Sleep Med. \textbf{3}, 121 (2007).
\bibitem{aasm2023} M.~M. Troester, S.~F. Quan, R.~B. Berry, \textit{et al.}; for the American Academy of Sleep Medicine, \textit{The AASM Manual for the Scoring of Sleep and Associated Events: Rules, Terminology and Technical Specifications, Version 3} (American Academy of Sleep Medicine, Darien, IL, 2023).
\bibitem{bak1987} P.~Bak, C.~Tang, and K.~Wiesenfeld, Phys. Rev. Lett. \textbf{59}, 381 (1987).
\bibitem{beggsplenz2003} J.~M. Beggs and D.~Plenz, J. Neurosci. \textbf{23}, 11167 (2003).
\bibitem{beggs2008} J.~M. Beggs, Philos. Trans. R. Soc. A \textbf{366}, 329 (2008).
\bibitem{disanto2018} S.~Di Santo, P.~Villegas, R.~Burioni, and M.~A. Mu\~noz, Proc. Natl. Acad. Sci. U.S.A. \textbf{115}, E1356 (2018).
\bibitem{iyer2018} K.~K. Iyer, Front. Neurosci. \textbf{12}, 948 (2018).
\bibitem{pearlmutter2009} B.~A. Pearlmutter and C.~Houghton, Neural Comput. \textbf{21}, 1622 (2009).
\bibitem{landau1937} L.~D. Landau, Zh. Eksp. Teor. Fiz. \textbf{7}, 19 (1937).
\bibitem{ginzburg1950} V.~L. Ginzburg and L.~D. Landau, Zh. Eksp. Teor. Fiz. \textbf{20}, 1064 (1950).
\bibitem{steynross2004} M.~L. Steyn-Ross, D.~A. Steyn-Ross, and J.~W. Sleigh, Prog. Biophys. Mol. Biol. \textbf{85}, 369 (2004).
\bibitem{steynross2005} M.~L. Steyn-Ross, D.~A. Steyn-Ross, J.~W. Sleigh, M.~T. Wilson, I.~P. Gillies, and J.~J. Wright, J. Biol. Phys. \textbf{31}, 547 (2005).
\bibitem{steynross2013} M.~L. Steyn-Ross, D.~A. Steyn-Ross, and J.~W. Sleigh, Phys. Rev. X \textbf{3}, 021005 (2013).
\bibitem{li2025} J.~Li, A.~Ilina, R.~Peach, T.~Wei, E.~Rhodes, V.~Jaramillo, I.~R. Violante, M.~Barahona, D.-J. Dijk, and N.~Grossman, Nat. Neurosci. \textbf{28}, 2515 (2025).
\bibitem{demooij2020} S.~M.~M. de Mooij, T.~F. Blanken, R.~P.~P.~P. Grasman, J.~R. Ramautar, E.~J.~W. Van Someren, and H.~L.~J. van der Maas, Comput. Methods Programs Biomed. \textbf{193}, 105448 (2020).
\bibitem{stevner2019} A.~B.~A. Stevner, D.~Vidaurre, J.~Cabral, \textit{et al.}, Nat. Commun. \textbf{10}, 1035 (2019).
\bibitem{sanzperl2020} Y.~S. Perl, H.~Bocaccio, I.~P\'erez-Ipi\~na, F.~Zamberl\'an, J.~Piccinini, H.~Laufs, M.~Kringelbach, G.~Deco, and E.~Tagliazucchi, Phys. Rev. Lett. \textbf{125}, 238101 (2020).
\bibitem{borbely1982} A.~A. Borb\'ely, Hum. Neurobiol. \textbf{1}, 195 (1982).
\bibitem{saper2001} C.~B. Saper, T.~C. Chou, and T.~E. Scammell, Trends Neurosci. \textbf{24}, 726 (2001).
\bibitem{saper2005} C.~B. Saper, T.~E. Scammell, and J.~Lu, Nature \textbf{437}, 1257 (2005).
\bibitem{lu2006} J.~Lu, D.~Sherman, M.~Devor, and C.~B. Saper, Nature \textbf{441}, 589 (2006).
\bibitem{robinson2011} P.~A. Robinson, A.~J.~K. Phillips, B.~D. Fulcher, M.~Puckeridge, and J.~A. Roberts, Philos. Trans. R. Soc. A \textbf{369}, 3840 (2011).
\bibitem{scheffer2009} M.~Scheffer, J.~Bascompte, W.~A. Brock, \textit{et al.}, Nature \textbf{461}, 53 (2009).
\bibitem{meisel2017} C.~Meisel, A.~Klaus, V.~V. Vyazovskiy, and D.~Plenz, J. Neurosci. \textbf{37}, 10114 (2017).
\bibitem{priesemann2013} V.~Priesemann, M.~Valderrama, M.~Wibral, and M.~Le Van Quyen, PLoS Comput. Biol. \textbf{9}, e1002985 (2013).
\bibitem{buendia2020} V.~Buend\'ia, S.~di Santo, P.~Villegas, R.~Burioni, and M.~A. Mu\~noz, Phys. Rev. Res. \textbf{2}, 013318 (2020).
\bibitem{borbely2016} A.~A. Borb\'ely, S.~Daan, A.~Wirz-Justice, and T.~Deboer, J. Sleep Res. \textbf{25}, 131 (2016).
\bibitem{hu2026} Z.~Hu, M.~Aravind, X.~Lei, J.~N. Kutz, and J.-J. Aucouturier, PLOS Comput. Biol. \textbf{22}, e1014246 (2026).
\bibitem{cash2009} S.~S. Cash, E.~Halgren, N.~Dehghani, \textit{et al.}, Science \textbf{324}, 1084 (2009).
\bibitem{degennaro2003} L.~De Gennaro and M.~Ferrara, Sleep Med. Rev. \textbf{7}, 423 (2003).
\bibitem{fernandez2020} L.~M.~J. Fernandez and A.~L\"uthi, Physiol. Rev. \textbf{100}, 805 (2020).
\bibitem{huber2004} R.~Huber, M.~F. Ghilardi, M.~Massimini, and G.~Tononi, Nature \textbf{430}, 78 (2004).
\bibitem{massimini2004} M.~Massimini, R.~Huber, F.~Ferrarelli, S.~Hill, and G.~Tononi, J. Neurosci. \textbf{24}, 6862 (2004).
\bibitem{nir2011} Y.~Nir, R.~J. Staba, T.~Andrillon, \textit{et al.}, Neuron \textbf{70}, 153 (2011).
\bibitem{vyazovskiy2011} V.~V. Vyazovskiy, U.~Olcese, E.~C. Hanlon, Y.~Nir, C.~Cirelli, and G.~Tononi, Nature \textbf{472}, 443 (2011).
\bibitem{achermann1997} P.~Achermann and A.~A. Borb\'ely, Neuroscience \textbf{81}, 213 (1997).
\bibitem{lempel1976} A.~Lempel and J.~Ziv, IEEE Trans. Inf. Theory \textbf{22}, 75 (1976).
\bibitem{costa2002} M.~Costa, A.~L. Goldberger, and C.-K. Peng, Phys. Rev. Lett. \textbf{89}, 068102 (2002).
\bibitem{iliff2012} J.~J. Iliff, M.~Wang, Y.~Liao, \textit{et al.}, Sci. Transl. Med. \textbf{4}, 147ra111 (2012).
\bibitem{xie2013} L.~Xie, H.~Kang, Q.~Xu, \textit{et al.}, Science \textbf{342}, 373 (2013).
\bibitem{hablitz2019} L.~M. Hablitz, H.~S. Vinitsky, Q.~Sun, \textit{et al.}, Sci. Adv. \textbf{5}, eaav5447 (2019).
\bibitem{fultz2019} N.~E. Fultz, G.~Bonmassar, K.~Setsompop, \textit{et al.}, Science \textbf{366}, 628 (2019).
\bibitem{dagum2026} P.~Dagum, D.~L. Elbert, L.~Giovangrandi, \textit{et al.}, Nat. Commun. \textbf{17}, 715 (2026).
\bibitem{miao2024} A.~Miao, T.~Luo, B.~Hsieh, \textit{et al.}, Nat. Neurosci. \textbf{27}, 1046 (2024).
\bibitem{hauglund2025} N.~L. Hauglund, M.~Andersen, K.~Tokarska, \textit{et al.}, Cell \textbf{188}, 606 (2025).
\bibitem{tononi2006} G.~Tononi and C.~Cirelli, Sleep Med. Rev. \textbf{10}, 49 (2006).
\bibitem{scammell2017} T.~E. Scammell, E.~Arrigoni, and J.~O. Lipton, Neuron \textbf{93}, 747 (2017).
\bibitem{passaro2025} A.~D. Passaro and A.~Poltorak, bioRxiv 10.1101/2024.12.16.628645 (2025).
\bibitem{kemp2000} B.~Kemp, A.~H. Zwinderman, B.~Tuk, H.~A.~C. Kamphuisen, and J.~J.~L. Oberye, IEEE Trans. Biomed. Eng. \textbf{47}, 1185 (2000).
\bibitem{donoghue2020} T.~Donoghue, M.~Haller, E.~J. Peterson, \textit{et al.}, Nat. Neurosci. \textbf{23}, 1655 (2020).
\bibitem{lendner2020} J.~D. Lendner, R.~F. Helfrich, B.~A. Mander, \textit{et al.}, eLife \textbf{9}, e55092 (2020).
\bibitem{peng1995} C.-K. Peng, S.~Havlin, H.~E. Stanley, and A.~L. Goldberger, Chaos \textbf{5}, 82 (1995).
\bibitem{hartigan1985} J.~A. Hartigan and P.~M. Hartigan, Ann. Stat. \textbf{13}, 70 (1985).
\bibitem{benjamini1995} Y.~Benjamini and Y.~Hochberg, J. R. Stat. Soc. B \textbf{57}, 289 (1995).
\bibitem{mander2017} B.~A. Mander, J.~R. Winer, and M.~P. Walker, Neuron \textbf{94}, 19 (2017).
\bibitem{hohenberg1977} P.~C. Hohenberg and B.~I. Halperin, Rev. Mod. Phys. \textbf{49}, 435 (1977).
\end{thebibliography}
\end{document}